\documentclass[12pt,draftcls,onecolumn]{IEEEtran}
\ifCLASSINFOpdf
  \usepackage[pdftex]{graphicx}
  \graphicspath{{../pdf/}{../jpeg/}}
  \DeclareGraphicsExtensions{.pdf,.jpeg,.png}
\else
  \usepackage[dvips]{graphicx}
  \graphicspath{{../eps/}}
  \DeclareGraphicsExtensions{.eps}
\fi\usepackage{graphicx}
\usepackage{graphics}
\usepackage{epsfig}\usepackage{epstopdf}

\usepackage{epstopdf}
\usepackage{stfloats}
\usepackage[cmex10]{amsmath}
\usepackage{algorithmic}
\usepackage{array}
\usepackage{mdwmath}
\usepackage{mdwtab}
\usepackage{graphicx}
\usepackage{subfigure}
\usepackage{color}
\usepackage{amsfonts,amssymb}
\usepackage{hyperref}

\begin{document}

\title{Secrecy Outage Analysis for Cooperative NOMA Systems with Relay Selection Scheme\thanks{Manuscript received.}}

\author{Hongjiang~Lei,
        Zixuan~Yang,
        Ki-Hong~Park,
        Imran Shafique Ansari\\
        Yongcai~Guo,
        Gaofeng~Pan,
        and~Mohamed-Slim~Alouini
\thanks{H. Lei and Z. Yang are with School of Communication and Information Engineering \& Chongqing Key Lab of
Mobile Communications Technology, Chongqing University of Posts and Telecommunications, Chongqing 400065, China
(e-mail: leihj@cqupt.edu.cn; cquptyzx@gmail.com).}
\thanks{I. S. Ansari is with the School of Engineering, University of Glasgow, Glasgow G12 8QQ, United Kingdom (e-mail: imran.ansari@glasgow.ac.uk).}
\thanks{Y. Guo is with the Key Laboratory of Optoelectronic Technology and Systems of the Education Ministry of China, Chongqing University, Chongqing 400044, China. (e-mail:ycguo@cqu.edu.cn).}
\thanks{G. Pan is with Chongqing Key Laboratory of Nonlinear Circuits and Intelligent Information Processing, Southwest University, Chongqing 400715, (e-mail: penngaofeng@qq.com).}
\thanks{K.-H. Park and M.-S. Alouini are with CEMSE Division, King Abdullah University of Science and Technology (KAUST), Thuwal 23955-6900, Saudi Arabia (e-mail: kihong.park@kaust.edu.sa; slim.alouini@kaust.edu.sa).}
}
\maketitle

\begin{abstract}
This paper considers the secrecy outage performance of a multiple-relay assisted non-orthogonal multiple access (NOMA) network over Nakagami-\emph{m} fading channels. Two slots are utilized to transmit signals from the base station to destination. At the first slot, the base station broadcasts the superposition signal of the two users to all decode-and-forward relays by message mapping strategy. Then the selected relay transmits superposition signal to the two users via power-domain NOMA technology. Three relay selection (RS) schemes, i.e., optimal single relay selection (OSRS) scheme, two-step single relay selection (TSRS) scheme, and optimal dual relay selection (ODRS) scheme, are proposed and the secrecy outage performance are analyzed.
As a benchmark, we also examine the secrecy outage performance of the NOMA systems with traditional multiple relays combining (TMRC) scheme in which all the relay that successfully decode signals from the source forward signals to the NOMA users with equal power. Considering the correlation between the secrecy capacity of two users and different secrecy requirement for two NOMA users, the closed-form expressions for the security outage probability (SOP) of the proposed OSRS, TSRS, and ODRS schemes along with the TMRC scheme are derived and validated via simulations. To get more insights, we also derive the closed-form expressions for the asymptotic SOP for all the schemes with fixed and dynamic power allocations. Furthermore, the secrecy diversity order (SDO) of cooperative NOMA systems is obtained. The results demonstrate that our proposed schemes can significantly enhance the secrecy performance compared to the TMRC scheme and that all the RS schemes with fixed power allocation obtain zero SDO and the OSRS scheme with dynamic power allocation obtains the same SDO as TMRC.
\end{abstract}



\section{Introduction}
\label{sec:introduction}
\subsection{Background and Related Works}
Non-orthogonal multiple access (NOMA) has been considered as one of the most promising technologies to deal with the shortage of bandwidth resources and enhance user fairness in the fifth generation (5G) mobile network \cite{DaiL2015Mag}-\cite{ZhouF2018WC}.
Compared to traditional orthogonal multiple access schemes, NOMA systems can obtain superior merits, such as high bandwidth efficiency, better user fairness, ultra-connectivity, and high flexibility. After receiving the superposition signal transmitted by the source node, the stronger user in NOMA systems firstly decodes the signal sent to the weaker user, cancels it from the received signals, and then decodes the signals for itself.

Cooperative communication is a particularly attractive technique that not only extends the network's coverage but also enhances the system performance when diversity technology is utilized at the destinations. Since the stronger users in NOMA systems always firstly decode the messages for the weaker users, therefore the stronger users can be utilized as relays to improve the performance of those weaker users with poor channel conditions \cite{Ding2015CL, LiuY2016JSACSWIPT}. Ding \emph{et al.} analyzed the outage probability (OP), diversity order of cooperative NOMA systems, and proposed an approach based on user pairing to reduce system complexity in \cite{Ding2015CL}.
Furthermore, a new cooperative simultaneous wireless information and power transfer NOMA protocol was proposed and the closed-form expressions for the OP and system throughput were derived in \cite{LiuY2016JSACSWIPT}.
Recently, many literatures focused on the cooperative NOMA systems with dedicated relay nodes. For example, a dedicated amplify-and-forward (AF) relay with multiple antennas was utilized in the cooperative NOMA systems and the closed-form expressions for the exact and lower bound of OP were derived in \cite{MenJ2015CL, MenJ2017TVT, XiaoY2018CL}. Moreover, a novel power allocation (PA) scheme for dual-hop relaying NOMA systems along with its ergodic sum-rate and OP were investigated in \cite{Duan2017Access}.

Full-duplex (FD) relay also has been utilized in cooperative NOMA systems to obtain higher spectral efficiency and better performance \cite{ZhongC2016CL}. For example, the stronger user working in the FD mode was utilized to forward signals to the weaker user in \cite{ZhangZ2016TVT} and \cite{LiuG2018TWC}. The closed-form expression for OP was obtained and the results showed that FD relays outperform half-duplex (HD) relays.
Liu \emph{et al.} analyzed the PA problems for HD and FD cooperative NOMA systems thereby obtaining the closed-form expression for the optimal power allocation policies in \cite{LiuG2018TWC}.
Since the weaker user decodes its signals by directly treating signals for the stronger user as interference, then the performance of the weaker users will become the bottleneck of NOMA systems in many scenarios. The dedicated FD relay was utilized to improve the performance of the weaker users in \cite{ChuTMC2018CL}, \cite{KaderMF2018TVT}. The performance of FD cooperative NOMA systems over Nakagami-$m$ channels was investigated in \cite{ChuTMC2018CL} and the analytical expressions for OP and ergodic rate were derived.
In \cite{KaderMF2018TVT}, a sharing FD relay was utilized in NOMA systems with two source-destination pairs and the ergodic rate, OP, and outage capacity were investigated while both perfect and imperfect self-interference cancellation (SIC) schemes were considered, respectively.

Relay selection (RS) technique is an effective scheme in making full use of space diversity with low implementation complexity and it can straightforwardly improve the spectral efficiency of cooperative systems \cite{ZouY2013JSAC, LeiH2017TCCN}.
Considering the quality of service (QoS) requirements for the two users are different, a two-stage single-relay-selection strategy and dual-relay-selection with fixed power allocation (FPA) was studied in \cite{Ding2016WCL} and \cite{ZhaoJ2018access}, respectively, in order to obtain the OP. The results showed that the two-stage strategy achieves a better OP and diversity gain.
Combining DF and AF relaying, Yang \emph{et al.} extended the two-stage RS scheme with dynamic power allocation (DPA) and derived the exact and asymptotic analytical expressions for OP in \cite{YangZ2017TVT}.
In \cite{XuP2018TVT}, considering whether the relay can correctly decode two users' messages, two optimal RS schemes were proposed for cooperative NOMA systems with FPA and DPA at the relays, respectively.
Yue \emph{et al.}, in \cite{YueX2018TCOM}, studied the performance of cooperative NOMA systems with two RS schemes in which the relay operates in either FD or HD mode. Modeling the spatial topology of relays with homogeneous  poisson point process (PPP), the performance of cooperative NOMA systems with two-stage RS scheme was analyzed in \cite{ChenJ2018Access} and a closed-form approximation for the OP was obtained.

To deal with the security issues followed by the explosive increase in cellular data, physical layer security (PLS) is emerging as one of the most promising ways to ensure secure communication basing on the time varying nature of the wireless channels \cite{Zou2016Proc}, \cite{WangHM2016Book}.
A new optimal PA strategy was proposed to maximize the secrecy sum rate of NOMA systems and the results confirmed that a significant secrecy performance for NOMA systems was obtained in \cite{ZhangY2016CL}.
In \cite{LiuY2017TWC}, the security performance of a NOMA-based large-scale network is considered, where both the NOMA users and eavesdroppers were modeled by homogeneous PPPs.
In \cite{Lei2017Access}, we investigated the secrecy outage performance of a multiple-input single-output (MISO) NOMA systems with transmit antenna selection (TAS) schemes and the closed-form expressions for secrecy outage probability (SOP) were derived. The results demonstrated that the proposed DPA scheme can achieve non-zero secrecy diversity order (SDO) for all the TAS schemes. The secrecy outage performance of multiple-input multiple-output (MIMO) NOMA systems with multiple legitimate and illegitimate receivers was studied in \cite{Lei2018TVT} and the closed-form expressions for SOP were subsequently obtained.
Feng \emph{et al.} proposed in \cite{FengY2018arXiv} a new PA to maximize the secrecy rate of the stronger user while guaranteeing the non-secure transmission rate requirement to the weaker user, but the secrecy performance of the weaker user was not considered.
Secrecy beamforming schemes were proposed for MISO-NOMA systems, cognitive MISO-NOMA systems, and MIMO-NOMA systems in \cite{LvL2018TVT}, \cite{ZhouF2018JSAC}, and \cite{ZhuZ2018WCL}, respectively.
The secrecy performance of a NOMA system with multiple eavesdroppers was investigated while zero-forcing and minimum mean-square error decoding schemes were utilized on the legitimate destinations in \cite{JiangK2018Access}.
A new joint subcarrier (SC) assignment and PA scheme was proposed to improve the security of the two-way relay NOMA systems in \cite{ZhangH2018JSAC}.

The secrecy performance of a cooperative NOMA system with a dedicated AF/DF relay was investigated and the closed-form expressions for the SOP were obtained in \cite{Chen2018TVT}.
In some scenarios, the relay in cooperative NOMA systems is untrusted and curious to decode the users' messages. Two new HD nonorthogonal AF schemes were proposed for cooperative NOMA systems with an untrusted relay and the secrecy rate maximization-based PA scheme was discussed in \cite{JiangW2018TVT}.
Arafa \emph{et al.} proposed two new relaying schemes to resist the untrusted relay and the secrecy performance of cooperative NOMA systems with new relaying schemes was analyzed and compared in \cite{Arafa2018arXiv}.
Feng \emph{et al.} proposed an artificial-noise (AN) aided scheme to build up secure cooperative NOMA systems with a FD relay and the closed-form expression for the SOP was derived in \cite{FengY2017GC}.
The secrecy performance of cooperative NOMA systems in which the stronger user behaves as a FD relay was investigated in \cite{LiuC2018ICC} and the SOP was analyzed with an assumption of the imperfect SIC.
The security-reliability tradeoff for both cooperative and non-cooperative NOMA schemes was analyzed and analytical expressions for SOP were derived in \cite{LiB2018TCOM}.

\subsection{Contributions}
In many works, such as \cite{LiuY2017TWC}, \cite{Lei2018TVT}, \cite{Chen2018TVT}, the worst-case scenario was considered, in which it is assumed that the eavesdroppers have powerful detection capabilities and they can extract the message from the signals transmitted from the resource. Under this assumption, one can realize that the secrecy capacity of both/all the users in NOMA systems are not independent \cite{Lei2018TVT}.
The major contributions of this paper are as follows,
\begin{enumerate}
\item In this work, we analyze a cooperative NOMA system with multiple relays and an eavesdropper employing RS schemes to enhance the secrecy performance. Three RS (OSRS, TSRS, and ODRS) schemes are proposed. For the purpose of comparison, we also investigate the secrecy performance of the NOMA systems with traditional multiple relays combining (TMRC) scheme in which all the relays that successfully decode signals from the source forward signals to the NOMA users with equal power. Considering the correlation between the secrecy capacity of two users, the closed-form expressions for the SOP under different RS schemes are derived and validated via simulations.

\item Different secrecy QoS is considered in our work, i.e., different secrecy threshold rates are requested for the two NOMA users. Although the weaker user cannot perform SIC, either the stronger or the weaker user will be the bottleneck of the cooperative NOMA system. Our results demonstrate that this depends on the power allocation parameters and secrecy rate thresholds.

\item To obtain more insights, we also derive the closed-form expressions for the asymptotic SOP under different RS schemes with FPA and DPA. Furthermore, the SDO of cooperative NOMA systems are derived. The results demonstrate that all the RS schemes with FPA obtain zero SDO and the OSRS scheme achieves the same SDO as TMRC scheme.

\item Differing from \cite{LiuY2017TWC}, \cite{Lei2018TVT}, \cite{Chen2018TVT}, wherein it has been assumed that the secrecy capacity for the stronger and weaker users is independent, correlation between the secrecy capacity of the two users and different secrecy requirement for the two users is considered in this work, which is more practical.
\end{enumerate}

\subsection{Organization}
This paper is organized as follows, in Section \ref{sec:SystemModel}, the system model and RS schemes are introduced. In Section \ref{sec:SOP}, we analyze the security outage performance of the proposed system.
The asymptotic SOP for cooperative NOMA systems with FPA and DPA are derived in Sections \ref{sec:SOPAsy} and \ref{sec:SOPAsyDPA}, respectively.
Numerical and simulation results are presented in Section \ref{sec:RESULTS} to demonstrate the security performance of the system. Finally, Section \ref{sec:Conclusion} concludes the paper.

\section{System Model}
\label{sec:SystemModel}

\begin{figure}[!t]
\centering
\includegraphics[width = 3.1in]{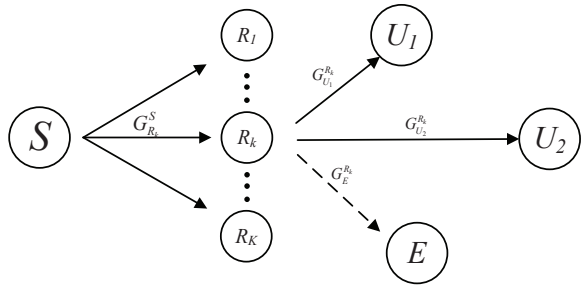}
\caption{System model demonstrating a cooperative NOMA system with one resource ($S$), multiple relays ($R_k$, $k = 1, \cdots ,K$), two legitimate user ($U_1$ and $U_2$), and an illegitimate eavesdropper ($E$).}
\label{figmodel}
\end{figure}
As shown in Fig. \ref{figmodel}, we consider a cooperative downlink NOMA system that consists of a base station ($S$), $K\left( {K \ge 1} \right)$ DF HD relays ($R_k$, $k = 1, \cdots ,K$), and two users ($U_1$ and $U_2$).
An eavesdropper ($E$) wants to wiretap the information through decoding the received signals.
It is assumed that all nodes are equipped with a single antenna and the direct link between $S$ and both users are unavailable due to deep fading.
The communication links between $S$ and all the receivers ($U_1$, $U_2$, and $E$) are relayed by $R_k$.

The probability density function (PDF) and the cumulative distribution function (CDF) of the channel power gains of link between the source node $\hbar$ and the destination node $\mathchar'26\mkern-10mu\lambda $ can be expressed by
\begin{equation}
{F_{{G_{\mathchar'26\mkern-10mu\lambda }^\hbar}}}\left( x \right) = 1 - {e^{ - \frac{{{m_{\mathchar'26\mkern-10mu\lambda }^\hbar}}}{{{\Omega _{\mathchar'26\mkern-10mu\lambda }^\hbar}}}x}}\sum\limits_{k = 0}^{{m_{\mathchar'26\mkern-10mu\lambda }^\hbar} - 1} {{{\left( {\frac{{{m_{\mathchar'26\mkern-10mu\lambda }^\hbar}}}{{{\Omega _{\mathchar'26\mkern-10mu\lambda }^\hbar}}}} \right)}^k}\frac{{{x^k}}}{{k!}}},
\label{CDFsingle}
\end{equation}
\begin{equation}
{f_{{G_{\mathchar'26\mkern-10mu\lambda }^\hbar}}}\left( x \right) = \frac{{{x^{{m_{\mathchar'26\mkern-10mu\lambda }^\hbar} - 1}}}}{{\Gamma \left( {{m_{\mathchar'26\mkern-10mu\lambda }^\hbar}} \right)}}{\left( {\frac{{{m_{\mathchar'26\mkern-10mu\lambda }^\hbar}}}{{{\Omega _{\mathchar'26\mkern-10mu\lambda }^\hbar}}}} \right)^{{m_{\mathchar'26\mkern-10mu\lambda }^\hbar}}}{e^{ - \frac{{{m_{\mathchar'26\mkern-10mu\lambda }^\hbar}}}{{{\Omega _{\mathchar'26\mkern-10mu\lambda }^\hbar}}}x}},
\label{PDFsingle}
\end{equation}
where $\hbar \in \left\{ {S,{R_k}} \right\}$ signifies the transmitter, ${\mathchar'26\mkern-10mu\lambda } \in \left\{ {{R_k},{U_1},{U_2},E} \right\}$ denotes the receiver,
$\Gamma \left( { \cdot } \right)$ is Gamma function, as defined by (8.310.1) of \cite{Gradshteyn2014},
$m_{\mathchar'26\mkern-10mu\lambda }^\hbar$ and $\Omega _{\mathchar'26\mkern-10mu\lambda }^\hbar$ are the fading parameters and average channel power gains, respectively.

To make analysis simple, it is assumed that the links of the first hop are independent and identically distributed (i.i.d.), which means $m_{{R_k}}^S = {m_R}$, $\Omega _{{R_k}}^S = {\Omega _R}$. The same assumption is made to the links between $R_k$ and each receiver in the second hop, which means
$m_{{U_1}}^{{R_k}} = m_{{U_2}}^{{R_k}} = {m_U}$, $\Omega _{{U_1}}^{{R_k}} = {\Omega _1}$, $\Omega _{{U_2}}^{{R_k}} = {\Omega _2}$, $m_E^{{R_k}} = {m_E}$, and $\Omega _E^{{R_k}} = {\Omega _E}$.

Two time slots are assumed to utilize to the communication between $S$ and the NOMA users.
In the first time slot, $S$ broadcasts the superposition signal to all relays by message mapping strategy, which is utilized and proved optimal to achieve the minimal common OP in \cite{XuP2018TVT}.
During the second time slot, a relay is selected using the RS scheme proposed below to send superposition signal to the two NOMA users. We assume all the relays are close enough such that the relative distance between the two users and the relay is determined.

To make the representation clear and simple, we use $G_{{U_i}}^{{R_k}} = G_i^k$.
It is also assumed that ${U_1}$ has a better channel condition than ${U_2}$ $\left( {G_1^k > G_2^k} \right)$, which is adopted in many NOMA studies, e.g., \cite{Liu2017Proc}, \cite{ZhongC2016CL}, \cite{ZhangZ2016TVT}, \cite{LiuG2018TWC}, \cite{XuP2018TVT}, \cite{Lei2017Access}.
Then the signal-to-interference-noise ratio (SINR) of the NOMA users can be written as
\begin{equation}
\gamma _1^k = {\alpha _1}\rho {G_1^k},
\label{snr1}
\end{equation}
\begin{equation}
\gamma _2^k = \frac{{{\alpha _2}\rho {G_2^k}}}{{{\alpha _1}\rho {G_2^k} + 1}},
\label{snr2}
\end{equation}
where ${\alpha _i}\,\left( {i = 1,2} \right)$ represents the power allocation coefficients at the $k$th relay, ${\alpha _1} + {\alpha _2} = 1$, ${\alpha _1} > {\alpha _2} $, and $\rho$ signifies the transmit signal to noise ratio (SNR).
\section{Secrecy Outage Probability Analysis}
\label{sec:SOP}
The set of relays that can correctly decode mixed signals from $S$ can be expressed as
\begin{equation}
{\Phi} \buildrel \Delta \over = \left\{ {k:1 \le k \le K,\frac{1}{2}{{\log }_2}\left( {1 + {\rho _S}G_{{R_k}}^S} \right) \ge R_1^{\rm{th}} + R_2^{\rm{th}}} \right\},
\label{fai0}
\end{equation}
where the factor $\frac{1}{2}$ arises from the fact that two slots are required to complete the data transmission,  $\rho_S$ denotes the transmit SNR at $S$, and $R_i^{\rm{th}}\;\left( {i = 1,2} \right)$ is the data rate threshold for $U_i$.
Thus the SOP of cooperative NOMA systems can be written as
\begin{equation}
{P_{\rm{out}}} = \sum\limits_{n = 0}^K {\Pr \left( {\left| {{\Phi}} \right| = n} \right){P_{{\Phi _n}}}},
\label{pout1}
\end{equation}
where ${P_{{\Phi _n}}} $ denotes the SOP under the condition that there are $n$ relays that correctly decode the mixed signals. One can easily have ${P_{{\Phi}_0}} = 1$.

It is assumed that all the links between the source and the relays are i.i.d., so we have
\begin{equation}
\begin{aligned}
\Pr \left\{ {\left| {{\Phi}} \right| = n} \right\} &= C_K^n{\left( {\Pr \left\{ {G_{{R_k}}^S \ge \eta } \right\}} \right)^n}{\left( {\Pr \left\{ {G_{{R_k}}^S < \eta } \right\}} \right)^{K - n}}\\
& = C_K^n{\chi ^n}{\left( {1 - \chi } \right)^{K - n}},
\label{pfain}
\end{aligned}
\end{equation}
where
$C_m^n = \frac{{m!}}{{n!\left( {m - n} \right)!}}$,
$\eta  = \frac{{{e^{2{R_1^{\rm{th}}} + 2{R_2^{\rm{th}}}}} - 1}}{{{\rho _S}}}$,
$\chi  = {e^{ - {\lambda _R}\eta }}\sum\limits_{k = 0}^{{m_R} - 1} {\frac{{\lambda _R^k{\eta ^k}}}{{k!}}} $, and
${\lambda _R} = \frac{{{m_R}}}{{{\Omega _R}}}$.

\subsection{Traditional Multiple Relays Combining Scheme}
As a benchmark, the traditional multiple relays combining (TMRC) scheme is presented in this subsection, where all the $n$ relays can successfully decode and forward the signal to both the users with equal power. Both the legitimate and illegitimate receivers combine their received signals with maximal ratio combining (MRC) scheme to maximize their SINR. To make fair comparison, it is assumed that the total transmit power at these relays is given by ${P_R}$. Then the ${P_{{\Phi _n}}}$ with this scheme can be expressed as
\begin{equation}
{P_{{\Phi _n}}} = \Pr \left\{ {{C_{s,1}^{{\rm{TMRC}}} < {R_1^s}\;{\rm{or}}\;C_{s,2}^{{\rm{TMRC}}} < {R_2^s}} \left| {\left| {{\Phi}} \right| = n} \right.} \right\},
\label{PfaiMR1}
\end{equation}
where $C_{s,i}^{{\rm{TMRC}}} = \ln \left( {\frac{{1 + \gamma _i^{{\rm{TMRC}}}}}{{1 + \gamma _{E,i}^{{\rm{TMRC}}}}}} \right)$ means the secrecy capacity of $U_i$, $i = 1,2$,
$\gamma _1^{{\rm{TMRC}}} = {\rho _1}{\alpha _1}G_1^{{\rm{TMRC}}}$,
$\gamma _2^{{\rm{TMRC}}} = \frac{{{\rho _1}{\alpha _2}G_2^{{\rm{TMRC}}}}}{{1 + {\rho _1}{\alpha _1}G_2^{{\rm{TMRC}}} }}$,
$G_i^{{\rm{TMRC}}} = \sum\limits_{k = 1}^n {G_i^k} $,
${\gamma _{E,i}^{{\rm{TMRC}}}}$ signifies the SNR at $E$ when $U_i$ is wiretapped,
$R_i^s$ is the secrecy rate threshold for $U_i$,
${\rho _1} = \frac{{{P_R}}}{{n{\sigma ^2}}}$,
and
${\sigma ^2}$ is the noise power.

The CDF of ${G_i^{{\rm{TMRC}}}}$ can be expressed as \cite{Lei2017TVTTAS}
\begin{equation}
\begin{aligned}
{F_{G_i^{{\rm{TMRC}}}}}\left( x \right) = 1 - {e^{ - {\lambda _i}x}}\sum\limits_{k = 0}^{{\tau _D} - 1} {\frac{{\lambda _i^k}}{{k!}}{x^k}},
\label{DCDFMR}
\end{aligned}
\end{equation}
where $i = 1, 2$, ${\lambda _i} = \frac{{{m_U}}}{{{\Omega _i}}}$, and ${\tau _U} = n{m_U}$.

As similar to \cite{LiuY2017TWC}, \cite{Lei2018TVT}, and \cite{Chen2018TVT}, we assume that $E$ has enough capabilities to detect multiuser data.
Then the SNR at $E$ can be expressed as
\begin{equation}
\gamma _{E,i}^{{\rm{TMRC}}} = {\rho _1}{\alpha _i}G_E^{{\rm{TMRC}}},
\end{equation}
where $i = 1, 2$ and $G_E^{{\rm{TMRC}}} = \sum\limits_{k = 1}^n {G_E^k} $.
Similarly, the PDF of $G_E^{{\rm{TMRC}}}$ can be expressed as \cite{Lei2017TVTTAS}
\begin{equation}
{f_{G_E^{{\rm{TMRC}}}}}\left( x \right) = {\beta _E}{x^{{\tau _E} - 1}}{e^{ - {\lambda _E}x}},
\label{EPDFMR}
\end{equation}
where ${\beta _E} = \frac{{\lambda _E^{{\tau _E}}}}{{\Gamma \left( {{\tau _E}} \right)}}$, ${\lambda _E}{ = }\frac{{{m_E}}}{{{\Omega _E}}}$, and ${\tau _E} = n{m_E}$.

We can express the secrecy connection probability (SCP) of $U_1$, which is the complementary of SOP, as
\begin{equation}
\Pr \left\{ {C_{s,1}^{{\rm{TMRC}}} > {R_1^s}} \right\} = \Pr \left\{ {G_1^{{\rm{TMRC}}} > {\delta _1}\left( {G_E^{{\rm{TMRC}}}} \right)} \right\},
\label{sopu1}
\end{equation}
where ${\delta _1}\left( x \right) = {b_1} + {\theta _1}x$, ${b_1} = \frac{{{\theta _1} - 1}}{{{\alpha _1}{\rho _1}}}$, and ${\theta _1} = {e^{2{R_1^s}}}$.

Similarly, we obtain the SCP of $U_2$ as 
\begin{equation}
\begin{aligned}
\Pr \left\{ {C_{s,2}^{{\rm{TMRC}}} > {R_1^s}} \right\} &= \Pr \left\{ {{\rho _1}G_2^{{\rm{TMRC}}}\left( {\underbrace {1 - {\theta _2}{\alpha _1} - {\theta _2}{\alpha _1}{\alpha _2}{\rho _1}G_E^{{\rm{TMRC}}}}_{ \buildrel \Delta \over = \Lambda }} \right) > {\theta _2} - 1 + {\theta _2}{\alpha _2}{\rho _1}G_E^{{\rm{TMRC}}}} \right\}\\
& = \Pr \left\{ {\Lambda  > 0,G_2^{{\rm{TMRC}}} > {\delta _2}\left( {G_E^{{\rm{TMRC}}}} \right)} \right\}\\
& = \Pr \left\{ {G_E^{{\rm{TMRC}}} < {a_1},G_2^{{\rm{TMRC}}} > {\delta _2}\left( {G_E^{{\rm{TMRC}}}} \right)} \right\},
\label{tmrsopu2}
\end{aligned}
\end{equation}
where ${a_1} = \frac{{1 - {\theta _2}{\alpha _1}}}{{{\rho _1}{\alpha _1}{\alpha _2}{\theta _2}}}$,
${\delta _2}\left( x \right) = {c_1} + \frac{{{\alpha _2}}}{{{d_1} - {e_1}{x}}}$,
${c_1} =  - \frac{1}{{{\alpha _1}{\rho _1}}}$, ${d_1} = {\alpha _1}\rho _1 \left( {1 - {\alpha _1}{\theta _2}} \right)$, ${e_1} = \rho _1^2\alpha _1^2{\alpha _2}{\theta _2}$, and
${\theta _2} = {e^{2{R_2^s}}}$.

\textbf{\emph{Remark 1}}:
One can easily find that secrecy outage would occur at $U_2$ when $1 - {\theta _2}{\alpha _1} < 0$. This means that in order to ensure a secure NOMA system ($U_2$), there is a constraint for the  power allocation coefficients, which is expressed as
\begin{equation}
{\alpha _1} \le \frac{1}{{{\theta _2}}} = {e^{ - 2R_2^s}}\;{\rm{or}}\;{\alpha _2} > 1 - \frac{1}{{{\theta _2}}} = 1 - {e^{ - 2R_2^s}}.
\end{equation}

Based on (\ref{sopu1}) and (\ref{tmrsopu2}), ${P_{{\Phi _n}}}$ in this case can be rewritten as 
\begin{equation}
\begin{aligned}
P_{{\Phi _n}}^{{\rm{TMRC}}} & = \Pr \left\{ {\left( {C_{s,1}^{{\rm{TMRC}}} < {R_1^s}\;{\rm{or}}\;C_{s,2}^{{\rm{TMRC}}} < {R_2^s}} \right)\left| {\left| {{\Phi}} \right| = n} \right.} \right\}\\
& = 1 - \Pr \left\{ {\left( {C_{s,1}^{{\rm{TMRC}}} > {R_1^s}\;,C_{s,2}^{{\rm{TMRC}}} > {R_2^s}} \right)\left| {\left| {{\Phi}} \right| = n} \right.} \right\}\\
&  = 1 - \Pr \left\{ {G_1^{{\rm{TMRC}}} > {\delta _1}\left( {G_E^{{\rm{TMRC}}}} \right),G_1^{{\rm{TMRC}}} > {\delta _2}\left( {{G_E^{{\rm{TMRC}}}}} \right),G_E^{{\rm{TMRC}}} < {a_1}} \right\}\\
& = 1 - \int_0^{{a_1}} {\Pr \left\{ {G_1^{{\rm{TMRC}}} > {\delta _1}\left( x \right)} \right\}\Pr \left\{ {G_2^{{\rm{TMRC}}} > {\delta _2}\left( x \right)} \right\}{f_{G_E^{{\rm{TMRC}}}}}\left( x \right)dx} \\
& = 1 - \int_0^{{a_1}} {\left( {1 - {F_{G_1^{{\rm{TMRC}}}}}\left( {{\delta _1}\left( x \right)} \right)} \right)\left( {1 - {F_{G_2^{{\rm{TMRC}}}}}\left( {{\delta _2}\left( x \right)} \right)} \right){f_{G_E^{{\rm{TMRC}}}}}\left( x \right)dx}.
\label{PfaiTMR}
\end{aligned}
\end{equation}

To facilitate the following analysis, we define
\begin{equation}
g\left( {a,b,c,r,q,f,h,k,j} \right) = \int_0^a {{x^{b - 1}}{e^{ - fx - \frac{h}{{1 - qx}}}}}{\left( {1 + cx} \right)^k}{\left( {1 + \frac{r}{{1 - qx}}} \right)^j}dx.
\label{g1}
\end{equation}

To the authors' best knowledge, it's very difficult to obtain the closed-form expression of $g\left( {a,b,c,r,q,f,h,k,j} \right)$. Here by making use of Gaussian-Chebyshev quadrature from eq. (25.4.30) of \cite{Abramowitz1972}, we obtain
\begin{equation}
\begin{aligned}
g\left( {a,b,c,r,q,f,h,k,j} \right) = {\left( {\frac{a}{2}} \right)^b}\sum\limits_{i = 1}^N {{w_i}s_i^{b - 1}{e^{ - \frac{{af{s_i}}}{2} - \frac{{2h}}{{2 - aq{s_i}}}}}}{\left( {1 + \frac{{ac{s_i}}}{2}} \right)^k}{\left( {1 + \frac{{2r}}{{2 - aq{s_i}}}} \right)^j},
\label{g2}
\end{aligned}
\end{equation}
where $N$ is the number of terms, ${s_i} = {t_i} + 1$, ${t_i}$ is the $i$th zero of Legendre polynomials, ${w_i}$ is the Gaussian weight, which is given in Table (25.4) of \cite{Abramowitz1972}.

Substituting (\ref{DCDFMR}) and (\ref{EPDFMR}) into (\ref{PfaiTMR}) and with some simple algebraic manipulations, we have
\setcounter{equation}{17}
\begin{equation}
\begin{aligned}
P_{{\Phi _n}}^{{\rm{TMRC}}} = 1 - {\beta _E}{e^{ - {\lambda _1}{b_1} - {\lambda _2}{c_1}}}\sum\limits_{k = 0}^{{\tau _U} - 1} {\sum\limits_{j = 0}^{{\tau _U} - 1} {\frac{{\lambda _1^k\lambda _2^j{\Xi _1}}}{{k!j!}}} },
\label{PfaiMR3}
\end{aligned}
\end{equation}
where ${\Xi _1} = b_1^kc_1^jg\left( {{a_1},{\tau _E},\frac{{{\theta _1}}}{{{b_1}}},\frac{{{\alpha _2}}}{{{d_1}{c_1}}},\frac{{{e_1}}}{{{d_1}}},{\lambda _1}{\theta _1} + {\lambda _E},\frac{{{\lambda _2}{\alpha _2}}}{{{d_1}}}},k, j \right)$.

\subsection{Optimal Single Relay Selection Scheme}
In this subsection, we propose OSRS scheme to minimize the overall SOP of the proposed cooperative NOMA system. Based on (\ref{sopu1}), one can easily obtain the SOP for $U_1$ with $m$th relay as
\begin{equation}
\begin{aligned}
{P_{out,1}} &= 1 - \Pr \left\{ {G_1^m > {\delta _1}\left( {G_E^m} \right)} \right\}\\
& = \Pr \left\{ {G_1^m < {\delta _3}\left( {G_E^m} \right)} \right\}\\
& = \Pr \left\{ {\frac{{G_1^m}}{{{\delta _3}\left( {G_E^m} \right)}} < 1} \right\} ,
\end{aligned}
\end{equation}
where
${\delta _3}\left( x \right) = {b_2} + {\theta _1}x$,
${b_2} = \frac{{{\theta _1} - 1}}{{{\alpha _1}{\rho _2}}}$, and
${\rho _2} = \frac{{{P_R}}}{{{\sigma ^2}}}$.

Similarly, based on (\ref{tmrsopu2}), we can express SOP for $U_2$ as
\begin{equation}
\begin{aligned}
{P_{out,2}} &= 1 - \Pr \left\{ {G_E^m < {a_2},G_2^m > {\delta _4}\left( {G_E^m} \right)} \right\}\\
& = 1 - \left( {\Pr \left\{ {G_E^m < {a_2}} \right\} - \Pr \left\{ {G_E^m < {a_2},G_2^m < {\delta _4}\left( {G_E^m} \right)} \right\}} \right)\\
& = \Pr \left\{ {G_E^m < {a_2},G_2^m < {\delta _4}\left( {G_E^m} \right)} \right\} + 1 - \Pr \left\{ {G_E^m < {a_2}} \right\}\\
& = \Pr \left\{ {G_2^m < {\delta _4}\left( {G_E^m} \right)\left| {G_E^m < {a_2}} \right.} \right\}\Pr \left\{ {G_E^m < {a_2}} \right\} + \Pr \left\{ {G_E^m > {a_2}} \right\}\\
& = \Pr \left\{ {\frac{{G_2^m}}{{{\delta _4}\left( {G_E^m} \right)}} < 1\left| {G_E^m < {a_2}} \right.} \right\}\Pr \left\{ {G_E^m < {a_2}} \right\} + \Pr \left\{ {G_E^m > {a_2}} \right\},
\label{eqpout2}
\end{aligned}
\end{equation}
where ${a_2} = \frac{{1 - {\theta _2}{\alpha _1}}}{{{\theta _2}{\alpha _1}{\alpha _2}{\rho _2}}}$,
${\delta _4}\left( x \right) = {c_2} + \frac{{{\alpha _2}}}{{{d_2} - {e_2}x}}$,
${c_2} =  - \frac{1}{{{\alpha _1}{\rho _2}}}$,
${d_2} = {\alpha _1}{\rho _2} - {\theta _2}\alpha _1^2{\rho _2}$, and
${e_2} = {\theta _2}\alpha _1^2{\alpha _2}\rho _2^2$. One can observe that when ${G_E^m > {a_2}}$ the secrecy outage would occur at $U_2$, which means the cooperative NOMA system is not secure.

For the $m$th relay, we define
\setcounter{equation}{20}
\begin{equation}
{X_m} = \left\{ {\begin{array}{*{20}{c}}
{\min \left\{ {\frac{{G_1^m}}{{{\delta _3}\left( {G_E^m} \right)}},\frac{{G_2^m}}{{{\delta _4}\left( {G_E^m} \right)}}} \right\}},&{G_E^m < {a_2}}\\
0,&{G_E^m \ge {a_2}}
\end{array}} \right..
\label{Xm}
\end{equation}
Then to maximize the secrecy performance the relay is selected with the following criterion
\begin{equation}
{m^*} = \arg \mathop {\max }\limits_{m \in {\Phi}} \left( {{X_m}} \right).
\label{OSRS}
\end{equation}

\textbf{\emph{Remark 2}}:
It should be noted that the selection scheme in (\ref{OSRS}) is different from the max-min transmit antenna selection scheme (MMTAS) proposed in \cite{Lei2018TVT}, which is to maximize the minimum secrecy capacity of two users. Here the selection criterion is to minimize the SOP of the cooperative NOMA system.

\textbf{\emph{Remark 3}}:
There is another important difference between MMTAS and OSRS, i.e., it is assumed that $R_1^s = R_2^s$ in MMTAS. However, in the OSRS scheme, $R_1^s$ and $R_2^s$ can be different. MMTAS is a special case of OSRS when it is assumed that $R_1^s = R_2^s$ and the secrecy capacities of the two users are independent.

The SOP of the cooperative NOMA system conditioned on $\left| {{\Phi}} \right| = n$ can be written as

\begin{equation}
\begin{aligned}
P_{{\Phi _n}}^{{\rm{OSRS}}} &= {\left( {\Pr \left\{ {{X_m} < 1} \right\}} \right)^n}\\
& = {\left( {1 - \Pr \left\{ {{X_m} > 1} \right\}} \right)^n}\\
& = {\left( {1 - \Pr \left\{ {\min \left\{ {\frac{{G_1^m}}{{{\delta _3}\left( {G_E^m} \right)}},\frac{{G_2^m}}{{{\delta _4}\left( {G_E^m} \right)}}} \right\} \ge 1,G_E^m < {a_2}} \right\}} \right)^n}\\
& = {\left( {1 - \underbrace {\Pr \left\{ {G_1^m \ge {\delta _3}\left( {G_E^m} \right),G_2^m \ge {\delta _4}\left( {G_E^m} \right),G_E^m < {a_2}} \right\}}_{ \buildrel \Delta \over = {\Delta _1}}} \right)^n}.
\label{SOPORS}
\end{aligned}
\end{equation}

Based on (\ref{sopu1}) and (\ref{tmrsopu2}), we have
\begin{equation}
\begin{aligned}
{\Delta _1} &= \Pr \left\{ {G_1^m \ge {\delta _3}\left( {G_E^m} \right),G_1^m \ge {\delta _4}\left( {G_E^m} \right),G_E^m < {a_2}} \right\}\\
& = \int_0^{{a_2}} {\left( {1 - {F_{G_1^m}}\left( {{\delta _3}\left( x \right)} \right)} \right)\left( {1 - {F_{G_2^m}}\left( {{\delta _4}\left( x \right)} \right)} \right){f_{G_E^m}}\left( x \right)dx} \\
& = {\beta _E}{e^{ - {\lambda _1}{b_2} - {\lambda _2}{c_2}}}\sum\limits_{k = 0}^{{m_U} - 1} {\frac{{\lambda _1^k}}{{k!}}} \sum\limits_{j = 0}^{{m_U} - 1} {\frac{{\lambda _2^j}}{{j!}}{\Xi _2}},
\label{Delta1}
\end{aligned}
\end{equation}
where
${\Xi _2} = b_2^kc_2^jg\left( {{a_2},{m_E},\frac{{{\theta _1}}}{{{b_2}}},\frac{{{\alpha _2}}}{{{d_2}{c_2}}},\frac{{{e_2}}}{{{d_2}}},{\lambda _1}{\theta _1} + {\lambda _E},\frac{{{\lambda _2}{\alpha _2}}}{{{d_2}}}},k, j \right)$.

\subsection{Two-Step Single Relay Selection Scheme}
In some scenarios stated in \cite{Ding2016WCL, ZhaoJ2018access}, the QoS requirements for the two users are different. We can easily obtain the similar conclusion: the secrecy QoS for one user is higher than that of the other user. In this subsection, a new two-step RS (TSRS) scheme is proposed for such scenarios and the follwing two purposes will be realized  simultaneously. One is to ensure there is no secrecy outage for the user that has lower secrecy QoS requirement and the other is to serve the user with higher secrecy QoS requirement with a secrecy rate as large as possible.
The TSRS scheme is presented as follows.

In the first step, the following subset is built in the relays by focusing on $U_1$'s target secrecy rate
\begin{equation}
{\Psi} = \left\{ {i:i \in {\Phi},C_{s,1}^i > {R_1^s}} \right\},
\label{fai1}
\end{equation}
where $C_{s,1}^i = \frac{1}{2}\ln \left( {\frac{{1 + {\alpha _1}{\rho _2}G_1^i}}{{1 + {\alpha _1}{\rho _2}G_E^i}}} \right)$ signifying the secrecy capacity for $U_1$ from ${R_i}$ and $i \in {\Phi}$.

At the second step, the relay to maximize the secrecy capacity of $U_2$ is selected, i.e.,
\begin{equation}
{j^*} = \arg \mathop {\max }\limits_{j \in {\Psi}} \left( {C_{s,2}^j} \right),
\end{equation}
where $C_{s,2}^j = \frac{1}{2}\ln \left( {\frac{{1 + \frac{{{\alpha _2}{\rho _2}G_2^j}}{{{\alpha _1}{\rho _2}G_2^j + 1}}}}{{1 + {\alpha _2}{\rho _2}G_E^j}}} \right)$ signifies the secrecy capacity for $U_2$ from ${R_j}\,
(j \in \Psi)$.

The SOP with this scheme can be achieved as
\begin{equation}
\begin{aligned}
P_{{\Phi _n}}^{{\rm{TSRS}}} & = \sum\limits_{j = 1}^n {C_n^j\Pr \left\{ {\mathop {\min }\limits_{1 \le i \le j} \left\{ {C_{s,1}^i} \right\} > {R_1^s},\mathop {\max }\limits_{1 \le k \le n - j} \left\{ {C_{s,1}^k} \right\} < {R_1^s},\mathop {\max }\limits_{1 \le p \le j} \left\{ {C_{s,2}^p} \right\} < {R_2^s}} \right\}} \\
& = \sum\limits_{j = 1}^n {C_n^j\Pr \left\{ {\mathop {\min }\limits_{1 \le i \le j} \left\{ {C_{s,1}^i} \right\} > {R_1^s},\mathop {\max }\limits_{1 \le p \le j} \left\{ {C_{s,2}^p} \right\} < {R_2^s}} \right\}\Pr \left\{ {\mathop {\max }\limits_{1 \le k \le n - j} \left\{ {C_{s,1}^k} \right\} < {R_1^s}} \right\}} \\
& = \sum\limits_{j = 1}^n {C_n^j{{\left( {\underbrace {\Pr \left\{ {C_{s,1}^i > {R_1^s},C_{s,2}^i < {R_2^s}} \right\}}_{ \buildrel \Delta \over = {\Delta _2}}} \right)}^j}{{\left( {\underbrace {\Pr \left\{ {C_{s,1}^k < {R_1^s}} \right\}}_{ \buildrel \Delta \over = {\Delta _3}}} \right)}^{n - j}}} \\
& = {\left( {\Pr \left\{ {C_{s,1}^i > {R_1^s},C_{s,2}^i < {R_2^s}} \right\} + \Pr \left\{ {C_{s,1}^i < {R_1^s}} \right\}} \right)^n}\\
& = {\left( {\Pr \left\{ {C_{s,1}^i < {R_1^s}\,\, {\rm{or}} \,\, {{C_{s,1}^i < R_1^s}}} \right\}} \right)^n}\\
& = {\left( {1 - \Pr \left\{ {C_{s,1}^i > R_1^s,C_{s,2}^i > R_2^s} \right\}} \right)^n}.
\label{SOPTSRpart2}
\end{aligned}
\end{equation}

\textbf{\emph{Remark 4}}:
An interesting result can be observed from (\ref{SOPTSRpart2}) that the expression for the SOP under TSRS scheme is the same as that under OSRS.


\subsection{Optimal Dual Relay Selection Scheme}
In this subsection, a new RS scheme named ODRS scheme is proposed, in which one relay is selected to transmit signals to NOMA users among the ones that can decode the mixed signals and another relay is selected to transmit AN among the ones that can not decode the source signals. In order to deteriorate the SINR at $E$, we select the jamming relay with the following criterion as
\setcounter{equation}{27}
\begin{equation}
{k^*} = \arg \mathop {\max }\limits_{k \in \bar \Phi } \left( {G_E^k} \right),
\end{equation}
where ${\bar \Phi }$ means the complement of ${\Phi }$.
For the sake of a fair comparison with TMRC and OSRS schemes, the total transmit power of all the relays is constrained to $P_R$. It is assumed that ${\alpha _{\rm{J}}}$ $\left( {0 \le {\alpha _{\rm{J}}} < 1} \right)$ portion is utilized to transmit the jamming signals, then the SINR at $E$ can be written as
\begin{equation}
\gamma _{E,i}^J = \frac{{{\alpha _i}{\rho _3}G_E^k}}{{1 + {\rho _4}{H_E} }},
\label{H232}
\end{equation}
where $i = 1, 2$, ${\rho _3} = \frac{{\left( {1 - {\alpha _{\rm{J}}}} \right){P_R}}}{{{\sigma ^2}}}$,
${\rho _4} = \frac{{{\alpha _{\rm{J}}}{P_R}}}{{{\sigma ^2}}}$,
and
${H_E} = \mathop {\max }\limits_{k \in \bar \Phi } \left\{ {G_E^k} \right\}$.

We assume that both users are aware of the jamming signals, which means jamming signals do not influence the SINR at both users \cite{LiuY2017TWC, LvL2018TVT, ZhouF2018JSAC}.
It must be noted that the special case of $n = K$ or $\alpha _{\rm{J}} = 0$, ODRS scheme reduces to OSRS scheme.
Then the SOP with ODRS scheme is expressed as
\begin{equation}
P_{out}^{\rm{ODRS}} = \sum\limits_{n = 0}^{K - 1} {\Pr \left( {\left| \Phi  \right| = n} \right)P_{{\Phi _n}}^{\rm{J}}}  + P_{{\Phi _K}}^{{\rm{OSRS}}},
\label{poutRSJ}
\end{equation}
where $P_{{\Phi _K}}^{{\rm{OSRS}}}$ is given in (\ref{SOPORS}).

Similar to (\ref{SOPTSRpart2}), the SOP of the cooperative NOMA system  on condition that $\left| \Phi  \right| = n \,\,\left( {n < K} \right)$ can be expressed as 
\begin{equation}
\begin{aligned}
P_{{\Phi _n}}^{\rm{J}} &= {\left( 1 - \Pr \left\{ {C_{s,1}^k > R_1^s{\mkern 1mu} {\mkern 1mu} {\rm{,}}{\mkern 1mu} {\mkern 1mu} C_{s,2}^k > R_2^s} \right\}\right)^n}\\
& = {\left( 1 - \Pr \left\{ {\ln \left( {\frac{{1 + {\alpha _1}{\rho _3}G_1^k}}{{1 + \frac{{{\alpha _1}{\rho _3}G_E^k}}{{1 + {\rho _4}{H_E}}}}}} \right) > R_1^s,\ln \left( {\frac{{1 + \frac{{{\alpha _2}{\rho _3}G_2^k}}{{1 + {\alpha _1}{\rho _3}G_2^k}}}}{{1 + \frac{{{\alpha _2}{\rho _3}G_E^k}}{{1 + {\rho _4}{H_E}}}}}} \right) > R_2^s} \right\}\right)^n}\\
& = {\left( 1 - \underbrace {\Pr \left\{ {G_1^k > \ell  + {\theta _1}Y,G_2^k > w + \frac{{wu}}{{1 - vY}},{Y < \frac{1}{v}} } \right\}}_{{\Delta _4}}\right)^n},
\label{SOPRSJ}
\end{aligned}
\end{equation}
where $\ell  = \frac{{{\theta _1} - 1}}{{{\alpha _1}{\rho _3}}}$,
$w =  - \frac{1}{{{\alpha _1}{\rho _3}}}$,
$u = \frac{{{\alpha _2}}}{{{\alpha _1}{\theta _2} - 1}}$,
$v = \frac{{{\alpha _1}{\alpha _2}{\theta _2}{\rho _3}}}{{1 - {a_1}{\theta _2}}}$, and
$Y = \frac{{G_E^k}}{{1 + {\rho _4}{H_E} + 1}}$.

To achieve the analytical expression of $\Delta _4$, the PDF of ${H_E}$ and $Y$ are given in \textbf{\emph{Lemma 1}} and \textbf{\emph{Lemma 2}}, respectively.

%
%

\textbf{\emph{Lemma 1:}}
Given as $\left| {\bar \Phi } \right| = K - n\,\,\left( {n < K} \right)$, the PDF of ${H_E}$ is given by
\setcounter{equation}{31}
\begin{equation}
{f_{{H_E}}}\left( z \right) = \frac{{(K - n){\lambda _E^{{m_E}}}}}{{\Gamma \left( {{m_E}} \right)}}\sum\limits_{S_E} {A{z^{B + {m_E} - 1}}{e^{ - C{\lambda _E}z}}},
\label{pdfHE}
\end{equation}
where ${S_E} = \left\{ {\left( {{n_1}, \cdots ,{n_{{m_E} + 1}}} \right) \in {\Bbb{N}}\left. {\left| {\sum\limits_{p = 1}^{{m_E} + 1} {{n_p}}  = K - n - 1} \right.} \right|} \right\}$,
$A = \left( {\frac{{\left( {K - n - 1} \right)!}}{{\prod\limits_{q = 1}^{{m_E} + 1} {\left( {{n_q}} \right)!} }}} \right)\prod\limits_{p = 2}^{{m_E} + 1} {{{\left( { - \frac{{\lambda _E^{p - 2}}}{{\left( {p - 2} \right)!}}} \right)}^{{n_p}}}} $,
$B = \sum\limits_{p = 2}^{{m_E} + 1} {{n_p}\left( {p - 2} \right)} $,
and
$C = 1 + \sum\limits_{p = 2}^{{m_E} + 1} {{n_p}}$.

Proof : See Appendix A.

\textbf{\emph{Lemma 2:}}
The PDF of $Y$ is given by
\begin{equation}
\begin{aligned}
{f_Y}\left( y \right) = {\varphi _0}\sum\limits_{k = 0}^{{m_E} - 1} {\sum\limits_{{S_E}} {\sum\limits_{j = 0}^k {\frac{{\delta {e^{ - {\lambda _E}y}}}}{{{{\left( {{\rho _4}y + C} \right)}^{\varsigma  + 1}}}}} } }( {{\rho _4}{\lambda _E}{y^{k + 1}} + D{y^k} - Ck{y^{k - 1}}}),
\label{pdfY}
\end{aligned}
\end{equation}
where
${\varphi _0} = \frac{{\left( {K - n} \right)\lambda _E^{{m_E}}}}{{\Gamma \left( {{m_E}} \right)}}$,
$\varsigma = B + {m_E} + j$,
$\delta  = \frac{{C_k^jA\rho _4^j\lambda _E^{k - \varsigma }\left( {\varsigma  - 1} \right)!}}{{k!}}$,
and $D = C{\lambda _E} - {\rho _4}k + {\rho _4}\varsigma $.

Proof : See Appendix B.

Similar to (\ref{Delta1}), based on (\ref{sopu1}) and (\ref{tmrsopu2}), we obtain the closed-form expression of $\Delta _4$ as
\begin{equation}
\begin{aligned}
{\Delta _4} &= \Pr \left\{ {G_1^k > \ell  + {\theta _1}Y,G_2^k > w + \frac{{wu}}{{1 - vy}},Y < \frac{1}{v}} \right\}\\
&= \int_0^{\frac{1}{v}} {\Pr \left\{ {G_1^k > \ell  + {\theta _1}Y,G_2^k > w + \frac{{wu}}{{1 - vy}}} \right\}} {f_Y}\left( y \right)dy\\
&= \int_0^{\frac{1}{v}} {\left( {1 - {F_{G_1^k}}\left( {\ell  + {\theta _1}y} \right)} \right)} \left( {1 - {F_{G_2^k}}\left( {w + \frac{{wu}}{{1 - vy}}} \right)} \right){f_Y}\left( y \right)dy\\
&= {e^{ - {\lambda _1}\ell  - {\lambda _2}w}}{\varphi _0} \sum\limits_{p = 0}^{{m_U} - 1} {\sum\limits_{q = 0}^{{m_U} - 1} {\sum\limits_{k = 0}^{{m_E} - 1} {\sum\limits_{{S_E}} {\sum\limits_{j = 0}^k {\frac{{\delta \lambda _1^p\lambda _2^q}{\Xi_3}}{{p!q!}}} } } } },
\label{Delta4}
\end{aligned}
\end{equation}
where ${\Xi_3} = {{w^q}h\left( {\frac{1}{v},p,q,{\lambda _1}{\theta _1} + {\lambda _E},{\lambda _2}wu,u,v,\ell} \right)}$
and
\begin{equation*}
\begin{aligned}
h\left( {a,b,c,f,r,u,v,\ell } \right) = \int_0^a {{{\left( {\ell  + {\theta _1}y} \right)}^b}{{\left( {1 + \frac{u}{{1 - vy}}} \right)}^c}}\frac{{\left( {{\rho _4}{\lambda _E}{y^{k + 1}} + D{y^k} - Ck{y^{k - 1}}} \right)}}{{{{\left( {{\rho _4}y + C} \right)}^{\varsigma  + 1}}{e^{fy + \frac{r}{{1 - vy}}}}}}dy.
\label{h1}
\end{aligned}
\end{equation*}
Similar to (\ref{g2}), utilizing Gaussian-Chebyshev quadrature, we have
\begin{equation*}
\begin{aligned}
h\left( {a,b,c,f,r,u,v,\ell } \right) = \frac{a}{2}\sum\limits_{i = 1}^N {{w_i}{{\left( {\ell  + {\theta _1}{s_j}} \right)}^b}{{\left( {1 + \frac{u}{{1 - v{s_j}}}} \right)}^c}} \frac{{\left( {{\rho _4}{\lambda _E}s_j^{k + 1} + Ds_j^k - Cks_j^{k - 1}} \right)}}{{{{\left( {{\rho _4}{s_j} + C} \right)}^{\varsigma  + 1}}{e^{f{s_j} + \frac{r}{{1 - v{s_j}}}}}}},
\label{h2}
\end{aligned}
\end{equation*}
where ${s_j} = \frac{a}{2}\left( {{t_i} + 1} \right)$.

\section{Asymptotic Secrecy Outage Probability Analysis}
\label{sec:SOPAsy}
To obtain more insights, in this section we analyze the asymptotic SOP in the higher SINR regime. Similar to \cite{Lei2017Access} and \cite{Lei2018TVT}, we assume that ${\Omega _2} \to \infty $, ${\Omega _1} = {\varepsilon _1}{\Omega _2}\left( {{\varepsilon _1} > 1} \right)$, and ${\Omega _R} = {\varepsilon _2}{\Omega _2}$, where $\varepsilon _1$ and $\varepsilon _2$ are constant.

The asymptotic CDF of $G_v^{k}$ $\left( {v \in \left\{ {1,2} \right\}} \right)$ can be expressed as \cite{Lei2017TVTTAS}
\setcounter{equation}{34}
\begin{equation}
F_{{G_v^{k}}}^\infty \left( x \right) = {\varphi _v}{x^{{\tau _U}}} + {\cal O}\left( {{x^{{\tau _U}}}} \right),
\label{cdfwuqiong}
\end{equation}
where ${\varphi _1} = \frac{{m_U^{{\tau _U}}}}{{{{\left( {\varepsilon _1 {\Omega _2}} \right)}^{{\tau _U}}}\left( {{\tau _U}} \right)!}}$,
${\varphi _2} = \frac{{m_U^{{\tau _U}}}}{{\Omega _2^{{\tau _U}}\left( {{\tau _U}} \right)!}}$,
and $\mathcal{O}\left(  \cdot  \right)$ denotes higher order terms.
Then the asymptotic SOP of cooperative NOMA systems is obtained as
\begin{equation}
P_{\rm{out}}^\infty  = \sum\limits_{n = 0}^K {C_K^n\varphi _R^{K - n}{\eta ^{{m_R}\left( {K - n} \right)}}P_{{\Phi _n}}^\infty },
\label{sopInf}
\end{equation}
where ${\varphi _R} = \frac{{{m_R}^{{m_R}}}}{{{{\left( {{\varepsilon _2}{\Omega _2}} \right)}^{{m_R}}}\left( {{m_R}} \right)!}}$. The ${P_{{\Phi _n}}^\infty }$ for all the RS schemes is given as follows.

\subsection{Traditional Multiple Relays Combining Scheme}
Based on (\ref{PfaiTMR}), we have the closed-form for SOP with TMRC scheme as
\begin{equation}
\begin{aligned}
P_{{\Phi _n}}^{{\rm{TMRC}},\infty } &= 1 - \int_0^{{a_1}} {\left( {1 - F_{{G_1}}^\infty \left( {{b_1} + {\theta _1}x} \right)} \right)\left( {1 - F_{{G_2}}^\infty \left( {{c_1} + \frac{{{\alpha _2}}}{{{d_1} - {e_1}x}}} \right)} \right){f_{{G_E}}}\left( x \right)dx} \\
&= 1 - {F_{{G_E}}}\left( {{a_1}} \right) + {\varphi _1}{\beta _E}\sum\limits_{k = 0}^{{\tau _U}} {\frac{{C_{{\tau _U}}^k\theta _1^kb_1^{{\tau _U} - k}\Upsilon \left( {k + {\tau _E},{\lambda _E}{a_1}} \right)}}{{\lambda _E^{k + {\tau _E}}}}} \\
&+ {\varphi _2}{\beta _E}c_1^{{\tau _U}}g\left( {{a_1},{\tau _E},0,\frac{{{\alpha _2}}}{{{c_1}{d_1}}},\frac{{{e_1}}}{{{d_1}}},{\lambda _E},0,0,{\tau _U}} \right)\\
&- {\varphi _1}{\varphi _2}{\beta _E}b_1^{{\tau _U}}c_1^{{\tau _U}}g\left( {{a_1},{\tau _E},\frac{{{\theta _1}}}{{{b_1}}},\frac{{{\alpha _2}}}{{{c_1}{d_1}}},\frac{{{e_1}}}{{{d_1}}},{\lambda _E},0,{\tau _U},{\tau _U}} \right),
\label{PfaiMRInf}
\end{aligned}
\end{equation}
where $\Upsilon \left( { \cdot , \cdot } \right)$ is the lower incomplete Gamma function, as defined by (8.350.1) of \cite{Gradshteyn2014}.

\textbf{\emph{Remark 5}}:
From (\ref{PfaiMRInf}), one can easily obtain
$P_{{\Phi _n}}^{{\rm{TMRC}},\infty }  \approx  1 - {{F_{{G_E}}}}\left( {{a_1}} \right)$ when ${\varphi _i} \to 0 $ $\left( {i \in \left\{ {1,2} \right\}} \right)$ with ${\Omega _2} \to \infty $.

\subsection{Optimal Single Relay Selection Scheme}
Based on (\ref{SOPORS}), we obtain
\setcounter{equation}{37}
\begin{equation}
P_{{\Phi _n}}^{{\rm{OSRS,}}\infty } = {\left( {1 - \Delta _1 ^\infty } \right)^n},
\label{PfaiORSInf}
\end{equation}
where $\Delta _1^\infty$ is given as
\begin{equation}
\begin{aligned}
\Delta _1^\infty  &= \int_0^{{a_2}} {\left( {1 - F_{G_1^m}^\infty \left( {{\delta _3}\left( x \right)} \right)} \right)\left( {1 - F_{G_2^m}^\infty \left( {{\delta _4}\left( x \right)} \right)} \right){f_{G_E^m}}\left( x \right)dx} \\
&= {F_{{G_E}}}\left( {{a_2}} \right) - {\varphi _3}{\beta _E}\sum\limits_{k = 0}^{{m_U}} {\frac{{C_{{m_U}}^k\theta _1^kb_2^{{m_U} - k}\Upsilon \left( {k + {m_E},{\lambda _E}{a_2}} \right)}}{{\lambda _E^{k + {m_E}}}}} \\
&- {\varphi _4}{\beta _E}c_2^{{m_U}}g\left( {{a_2},{m_E},0,\frac{{{\alpha _2}}}{{{c_2}{d_2}}},\frac{{{e_2}}}{{{d_2}}},{\lambda _E},0,0,{m_U}} \right)\\
&+ {\varphi _3}{\varphi _4}{\beta _E}b_2^{{m_U}}c_2^{{m_U}}g\left( {{a_2},{m_E},\frac{{{\theta _1}}}{{{b_2}}},\frac{{{\alpha _2}}}{{{c_2}{d_2}}},\frac{{{e_2}}}{{{d_2}}},{\lambda _E},0,{m_U},{m_U}} \right),
\label{Delta1Inf}
\end{aligned}
\end{equation}
where ${\varphi _3} = \frac{{m_U^{{m_U}}}}{{{{\left( {{\varepsilon _1}{\Omega _2}} \right)}^{{m_U}}}\left( {{m_U}} \right)!}}$ and ${\varphi _4} = \frac{{m_U^{{m_U}}}}{{\Omega _2^{{m_U}}\left( {{m_U}} \right)!}}$. Similarly, we obtain $P_{{\Phi _n}}^{{\rm{OSRS,}}\infty } \approx {\left( {1 - {F_{{G_E}}}\left( {{a_2}} \right)} \right)^n}$ when ${\varphi _i} \to 0 $ $\left( {i \in \left\{ {1,2} \right\}} \right)$ with ${\Omega _2} \to \infty $.

\subsection{Optimal Dual Relay Selection Scheme}
Based on (\ref{poutRSJ}), we obtain the asymptotic SOP for cooperative NOMA systems with ODRS scheme as
\setcounter{equation}{39}
\begin{equation}
P_{{\rm{out}}}^{{\rm{ODRS}},\infty } = \sum\limits_{n = 0}^{K - 1} {C_K^n\varphi _R^{K - n}{\eta ^{{m_R}\left( {K - n} \right)}}P_{{\Phi _n}}^{{\rm{ODRS}},\infty }}  + P_{{\Phi _K}}^{{\rm{OSRS,}}\infty }.
\label{sopRSJInf}
\end{equation}
Based on (\ref{SOPRSJ}), we have
\begin{equation}
P_{{\Phi _n}}^{{\rm{ODRS,}}\infty } = {\left( {1 - \Delta _4 ^\infty } \right)^n},
\label{PfaiRSJInf}
\end{equation}
where  $\Delta _4^\infty$  was shown as
\begin{equation}
\begin{aligned}
\Delta _4^\infty &= \int_0^{\frac{1}{v}} {\left( {1 - F_{G_1^k}^\infty \left( {\ell  + {\theta _1}y} \right)} \right)} \left( {1 - F_{G_2^k}^\infty \left( {w + \frac{{wu}}{{1 - vy}}} \right)} \right){f_Y}\left( y \right)dy\\
&= 1 - {\varphi _0}\sum\limits_{k = 0}^{{m_E} - 1} {\sum\limits_{{S_E}} {\sum\limits_{j = 0}^k {\frac{{\delta {e^{ - \frac{{{\lambda _E}}}{v}}}}}{{{{\left( {C + \frac{{{\rho _4}}}{v}} \right)}^\varsigma }{v^k}}}} } }  - {\varphi _0}{\varphi _3}\sum\limits_{k = 0}^{{m_E} - 1} {\sum\limits_{{S_E}} {\sum\limits_{j = 0}^k {\delta h\left( {\frac{1}{v},{m_U},0,{\lambda _E},0,u,v,\ell } \right)} } } \\
&- {\varphi _0}{\varphi _4}{w^{{m_U}}}\sum\limits_{k = 0}^{{m_E} - 1} {\sum\limits_{{S_E}} {\sum\limits_{j = 0}^k \delta  } } h\left( {\frac{1}{v},0,{m_U},{\lambda _E},0,u,v,\ell } \right)\\
&+ {\varphi _0}{\varphi _3}{\varphi _4}{w^{{m_U}}}\sum\limits_{k = 0}^{{m_E} - 1} {\sum\limits_{{S_E}} {\sum\limits_{j = 0}^k \delta  } } h\left( {\frac{1}{v},{m_U},{m_U},{\lambda _E},0,u,v,\ell } \right).
\label{Delta4Inf}
\end{aligned}
\end{equation}

Similarly, we have
\begin{equation*}
P_{{\Phi _n}}^{{\rm{ODRS,}}\infty } \approx {\left( {{\varphi _0}\sum\limits_{k = 0}^{{m_E} - 1} {\sum\limits_{{S_E}} {\sum\limits_{j = 0}^k {\delta {v^{ - k}}{e^{ - \frac{{{\lambda _E}}}{v}}}{{\left( {C + \frac{{{\rho _4}}}{v}} \right)}^{ - \varsigma }}} } } } \right)^n},
\end{equation*}
when ${\varphi _i} \to 0 $ $\left( {i \in \left\{ {1,2} \right\}} \right)$ with ${\Omega _2} \to \infty $.

\textbf{\emph{Remark 6}}:
Based on the definition of SDO as ${G_d} =  - \mathop {\lim }\limits_{{\Omega _2} \to \infty } \frac{{\ln \left( {P_{\rm{out}}^\infty } \right)}}{{\ln \left( {{\Omega _2}} \right)}}$, we obtain the SDO for the three RS schemes as $G_d^{{\rm{TMRC}}} = G_d^{{\rm{OSRS}}} = G_d^{{\rm{ODRS}}} = 0$ because the first term is independent of $\Omega _2$ in (\ref{PfaiMRInf}), (\ref{Delta1Inf}), and (\ref{Delta4Inf}), respectively.

\section{Asymptotic Secrecy Outage Probability Analysis with Dynamic Power Allocation Scheme}
\label{sec:SOPAsyDPA}
Since there is a ceiling for the SINR of $U_2$, which equals to $\frac{{{\alpha _2}}}{{{\alpha _1}}}$, there is a floor for the SOP of NOMA systems when the transmit SNR increases. A new DPA scheme was proposed in \cite{Lei2017Access} and the PA coefficients are given as
\setcounter{equation}{42}
\begin{equation}
\left\{ {\begin{array}{*{20}{c}}
{{\alpha _1^{{\rm{DPA}}}} = \frac{1}{{1 + \mu \lambda _{2}^{ - \varpi  }}}}\\
{{\alpha _2^{{\rm{DPA}}}} = \frac{{\mu \lambda _{2}^{ - \varpi  }}}{{1 + \mu \lambda _{2}^{ - \varpi  }}}}
\end{array}} \right.,
\label{DPAP}
\end{equation}
where $0 < \varpi  < 1$ and $\mu  > 1$.
It was testified that non-zero SDO can be obtained due to $\frac{{\alpha _2^{{\rm{DPA}}}}}{{\alpha _1^{{\rm{DPA}}}}} = \mu \lambda _2^{ - \varpi } \to 0$ with ${\Omega _2} \to \infty $ in \cite{Lei2017Access}
and \cite{Lei2018TVT}. In this section,  the asymptotic SOP is analyzed while this DPA scheme is utilized.
Based on (\ref{DPAP}), we have
$a_i^{{\rm{DPA}}} = \frac{{{{\left( {1 + \mu \lambda _2^{ - \varpi  }} \right)}^2} - {\theta _2}\mu \lambda _2^{ - \varpi  } - {\theta _2}}}{{{\theta _2}\mu \lambda _2^{ - \varpi  }{\rho _i}}}$,
$b_i^{{\rm{DPA}}} = \frac{{{\theta _1} - 1}}{{{\rho _i}}}\left( {1 + \mu \lambda _2^{ - \varpi  }} \right)$,
$c_i^{{\rm{DPA}}} =  - \frac{{1 + \mu \lambda _2^{ - \varpi  }}}{{{\rho _i}}}$,
$d_i^{{\rm{DPA}}} = \frac{{\left( {1 + \mu \lambda _2^{ - \varpi  } - {\theta _2}} \right){\rho _i}}}{{{{\left( {1 + \mu \lambda _2^{ - \varpi  }} \right)}^2}}}$,
$e_i^{\rm{DPA}} = \frac{{{\theta _2}\mu \lambda _2^{ - \varpi  }\rho _i^2}}{{{{\left( {1 + \mu \lambda _2^{ - \varpi  }} \right)}^3}}}$, and $i = 1, 2$.

\subsection{Traditional Multiple Relays Combining Scheme with Dynamic Power Allocation Scheme}
Substitute (\ref{DPAP}) into (\ref{PfaiMRInf}), we achieve the closed-form expression for asymptotic SOP under TMRC-DPA scheme as 
\begin{equation}
\begin{aligned}
P_{{\Phi _n}}^{{\rm{TMRC}},\infty ,{\rm{DPA}}} &= 1 - {F_{{G_E}}}\left( {a_1^{{\rm{DPA}}}} \right) + {\varphi _1}{\beta _E}\sum\limits_{k = 0}^{{\tau _U}} {\frac{{C_{{\tau _U}}^k\theta _1^k{{\left( {b_1^{{\rm{DPA}}}} \right)}^{{\tau _U} - k}}\Upsilon \left( {k + {\tau _E},{\lambda _E}a_1^{{\rm{DPA}}}} \right)}}{{\lambda _E^{k + {\tau _E}}}}} \\
& + {\varphi _2}{\beta _E}{\left( {c_1^{{\rm{DPA}}}} \right)^{{\tau _U}}}g\left( {a_1^{{\rm{DPA}}},{\tau _E},0,\frac{{\alpha _2^{{\rm{DPA}}}}}{{c_1^{{\rm{DPA}}}d_1^{{\rm{DPA}}}}},\frac{{e_1^{{\rm{DPA}}}}}{{d_1^{{\rm{DPA}}}}},{\lambda _E},0,0,{\tau _U}} \right)\\
& - {\varphi _1}{\varphi _2}{\beta _E}{\left( {b_1^{{\rm{DPA}}}} \right)^{{\tau _U}}}{\left( {c_1^{{\rm{DPA}}}} \right)^{{\tau _U}}}g\left( {a_1^{{\rm{DPA}}},{\tau _E},\frac{{{\theta _1}}}{{b_1^{{\rm{DPA}}}}},\frac{{\alpha _2^{{\rm{DPA}}}}}{{c_1^{{\rm{DPA}}}d_1^{{\rm{DPA}}}}},\frac{{e_1^{{\rm{DPA}}}}}{{d_1^{{\rm{DPA}}}}},{\lambda _E},0,{\tau _U},{\tau _U}} \right).
\label{DPATMRPfain}
\end{aligned}
\end{equation}
Then the SDO of the cooperative NOMA systems in this case is obtained as
\setcounter{equation}{44}
\begin{equation}
\begin{aligned}
G_d^{{\rm{TMRC}},{\rm{DPA}}} & = \mathop {\min }\limits_{0 \le n \le K} \left\{ {{m_R}\left( {K - n} \right) + {\tau _U}\left( {1 - \varpi} \right)} \right\}\\
& = \mathop {\min }\limits_{0 \le n \le K} \left\{ {K{m_R} - n\left( {{m_R} - {m_U}\left( {1 - \varpi} \right)} \right)} \right\}\\
& = \left\{ {\begin{array}{*{20}{c}}
{K{m_U}\left( {1 - \varpi } \right)},&{{m_R} > {m_U}\left( {1 - \varpi } \right)}\\
{K{m_R}},&{{m_R} < {m_U}\left( {1 - \varpi } \right)}
\end{array}} \right.\\
& = K\min\left\{ {{m_U}\left( {1 - \varpi } \right),{m_R}} \right\}.
\label{GdTMRDPA}
\end{aligned}
\end{equation}

\setcounter{equation}{45}
\subsection{Optimal Single Relay Selection Scheme with Dynamic Power Allocation Scheme}
Similar to (\ref{PfaiORSInf}), we have
\begin{equation}
P_{{\Phi _n}}^{{\rm{OSRS,}}\infty {\rm{,DPA}}} = {\left( {1 - \Delta _1 ^{\infty ,{\rm{DPA}}}} \right)^n},
\end{equation}
where $\Delta _1^{\infty ,{\rm{DPA}}}$ is given by
\begin{equation}
\begin{aligned}
\Delta _1^{\infty ,{\rm{DPA}}} &= {F_E}\left( {a_2^{{\rm{DPA}}}} \right) - {\varphi _3}{\beta _E}\sum\limits_{k = 0}^{{m_U}} {\frac{{C_{{m_U}}^k\theta _1^k{{\left( {b_2^{{\rm{DPA}}}} \right)}^{{m_U} - k}}\Upsilon \left( {k + {m_E},{\lambda _E}a_2^{{\rm{DPA}}}} \right)}}{{\lambda _E^{k + {m_E}}}}} \\
&- {\varphi _4}{\beta _E}{\left( {c_2^{{\rm{DPA}}}} \right)^{{m_U}}}g\left( {a_2^{{\rm{DPA}}},{m_E},0,\frac{{\alpha _2^{{\rm{DPA}}}}}{{c_2^{{\rm{DPA}}}d_2^{{\rm{DPA}}}}},\frac{{e_2^{{\rm{DPA}}}}}{{d_2^{{\rm{DPA}}}}},{\lambda _E},0,0,{m_U}} \right)\\
&+ {\varphi _3}{\varphi _4}{\beta _E}{\left( {b_2^{{\rm{DPA}}}} \right)^{{m_U}}}{\left( {c_2^{{\rm{DPA}}}} \right)^{{m_U}}}g\left( {a_2^{{\rm{DPA}}},{m_E},\frac{{{\theta _1}}}{{b_2^{{\rm{DPA}}}}},\frac{{\alpha _2^{{\rm{DPA}}}}}{{c_2^{{\rm{DPA}}}d_2^{{\rm{DPA}}}}},\frac{{e_2^{{\rm{DPA}}}}}{{d_2^{{\rm{DPA}}}}},{\lambda _E},0,{m_U},{m_U}} \right).
\label{DPAdelta1_inf}
\end{aligned}
\end{equation}

Then the SDO of the cooperative NOMA systems in this case is obtained as
\begin{equation}
\begin{aligned}
G_d^{{\rm{OSRS}},{\rm{DPA}}} & = \mathop {\min }\limits_{0 \le n \le K} \left\{ {{m_R}\left( {K - n} \right) + {nm _D}\left( {1 - \varpi } \right)} \right\}\\
& = G_d^{{\rm{TMRC}},{\rm{DPA}}}.
\label{GdORSDPA}
\end{aligned}
\end{equation}

\textbf{\emph{Remark 7}}:
Based on (\ref{GdTMRDPA}) and (\ref{GdORSDPA}), one can observe the SDO of cooperative NOMA systems under both TMRC-DPA and OSRS-DPA schemes equals to the minimum SDO of dual hops. Moreover, it is a full SDO based on the results in \cite{Lei2017Access} and \cite{Lei2018TVT}.

%

\setcounter{equation}{48}
\subsection{Optimal Dual Relay Selection with Dynamic Power Allocation Scheme}
Similar to (\ref{sopRSJInf}), we write the asymptotic SOP for the cooperative NOMA system under ODRS-DPA scheme as
\begin{equation}
P_{{\rm{out}}}^{{\rm{ODRS}},\infty ,{\rm{DPA}}} = \sum\limits_{n = 0}^{K - 1} {C_K^n\varphi _R^{K - n}{\eta ^{{m_R}\left( {K - n} \right)}}P_{{\Phi _n}}^{{\rm{ODRS}},\infty ,{\rm{DPA}}}} + P_{{\Phi _K}}^{{\rm{OSRS}},\infty ,{\rm{DPA}}}.
\end{equation}
Similar to (\ref{PfaiRSJInf}), we have
\begin{equation}
P_{{\Phi _n}}^{{\rm{ODRS,}}\infty {\rm{,DPA}}} = {\left( {1 - \Delta _4 ^{\infty ,{\rm{DPA}}}} \right)^n},
\end{equation}
where ${\Delta _4^{\infty ,{\rm{DPA}}}}$ is shown as
\begin{equation}
\begin{aligned}
\Delta _4^{\infty ,{\rm{DPA}}} &= 1 - {\varphi _0} \sum\limits_{k = 0}^{{m_E} - 1} {\sum\limits_{{S_E}} {\sum\limits_{j = 0}^k {\delta {e^{ - \frac{{{\lambda _E}}}{{{v^{{\rm{DPA}}}}}}}}{{\left( {C + \frac{{{\rho _4}}}{v^{\rm{DPA}}}} \right)}^{ - \varsigma }}{{\left( {{v^{{\rm{DPA}}}}} \right)}^{ - k}}} } } \\
& - {\varphi _0} {\varphi _3}\sum\limits_{k = 0}^{{m_E} - 1} {\sum\limits_{{S_E}} {\sum\limits_{j = 0}^k {\delta h\left( {\frac{1}{{{v^{{\rm{DPA}}}}}},{m_U},0,{\lambda _E},0,{u^{{\rm{DPA}}}},{v^{{\rm{DPA}}}},{\ell ^{{\rm{DPA}}}}} \right)} } } \\
& - {\varphi _0} {\varphi _4}{\left( {{w^{{\rm{DPA}}}}} \right)^{{m_U}}}\sum\limits_{k = 0}^{{m_E} - 1} {\sum\limits_{{S_E}} {\sum\limits_{j = 0}^k \delta  } } h\left( {\frac{1}{{{v^{{\rm{DPA}}}}}},0,{m_U},{\lambda _E},0,{u^{{\rm{DPA}}}},{v^{{\rm{DPA}}}},{\ell ^{{\rm{DPA}}}}} \right)\\
& + {\varphi _0} {\varphi _3}{\varphi _4}{\left( {{w^{{\rm{DPA}}}}} \right)^{{m_U}}}\sum\limits_{k = 0}^{{m_E} - 1} {\sum\limits_{{S_E}} {\sum\limits_{j = 0}^k \delta  } } h\left( {\frac{1}{{{v^{{\rm{DPA}}}}}},{m_U},{m_U},{\lambda _E},0,{u^{{\rm{DPA}}}},{v^{{\rm{DPA}}}},{\ell ^{{\rm{DPA}}}}} \right),
\label{Delta4InfDPA}
\end{aligned}
\end{equation}
where ${w ^{{\rm{DPA}}}} =  - \frac{1}{{\alpha _1^{{\rm{DPA}}}{\rho _3}}}$,
${v^{{\rm{DPA}}}} = \frac{{\alpha _1^{{\rm{DPA}}}\alpha _2^{{\rm{DPA}}}{\theta _2}{\rho _3}}}{{1 - \alpha _1^{{\rm{DPA}}}{\theta _2}}}$,
${u^{{\rm{DPA}}}} = \frac{{\alpha _2^{{\rm{DPA}}}}}{{\alpha _1^{{\rm{DPA}}}{\theta _2} - 1}}$,
and
${\ell ^{{\rm{DPA}}}} = \frac{{{\theta _1} - 1}}{{\alpha _1^{{\rm{DPA}}}{\rho _3}}}$.

\setcounter{equation}{51}

Then the SDO of the cooperative NOMA systems in this case is obtained as
\begin{equation}
\begin{aligned}
G_d^{{\rm{ODRS}},{\rm{DPA}}} &= \min \left\{ {\mathop {\min }\limits_{0 \le n \le K - 1} \left\{ {{m_R}\left( {K - n} \right) + n{H_1}} \right\},K(1 - \varpi){m_D}} \right\}\\
& = \min \left\{ {\mathop {\min }\limits_{0 \le n \le K - 1} \left\{ {K{m_R} + n\left( {{H_1} - {m_R}} \right)} \right\},K(1 - \varpi){m_D}} \right\}\\
& = \left\{ {\begin{array}{*{20}{c}}
{\min \left\{ {K{m_D}\left( {1 - \varpi} \right) + {m_R} - {H_2},K{m_D}(1 - \varpi)} \right\},}&{{m_R} > {H_1}}\\
{\min \left\{ {K{m_R},K{m_D}(1 - \varpi)} \right\},}&{{m_R} < {H_1}}
\end{array}} \right.\\
& = \left\{ {\begin{array}{*{20}{c}}
{K{m_R},}&{{m_R} < {H_1}}\\
{K{m_D}\left( {1 - \varpi} \right) + {m_R} - {H_2},}&{{H_1} < {m_R} < {H_2}}\\
{K{m_D}(1 - \varpi),}&{{m_R} > {H_2}}
\end{array}} \right.\\
& = \min \left\{ {K{m_R},\left( {K - 1} \right)\left( {{m_D}\left( {1 - \varpi } \right) - \varpi } \right) + {m_R},K{m_D}(1 - \varpi )} \right\},
\label{GdRSJDPA}
\end{aligned}
\end{equation}
where ${H_1} = {m_D}\left( {1 - \varpi} \right) - \varpi$ and ${H_2} = {m_D}\left( {1 - \varpi} \right) + \varpi\left( {K - 1} \right)$.

\textbf{\emph{Remark 8}}:
Based on (\ref{GdRSJDPA}), one can realize that using a single relay to send jamming signals may reduce the SDO of cooperative NOMA systems in some conditions.

\begin{figure}[!t]
\centering
\includegraphics[width = 2.8in]{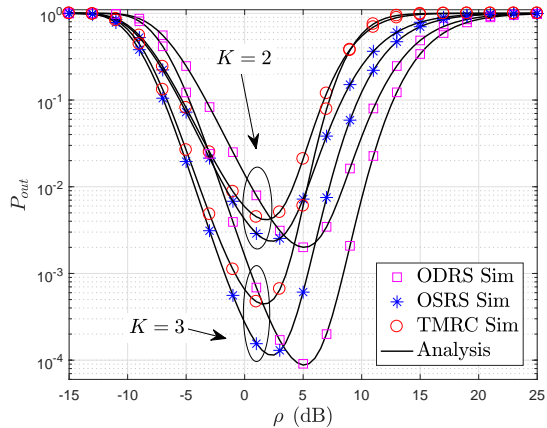}
\caption{SOP for various $K$ with ${R_1} = 0.1$, ${R_2} = 0.2$, ${m_U} = {m_E} = {m_R} = 2$, ${\alpha _{\rm{J}}} = 0.5$, ${\Omega _1} = 12$ dB, ${\Omega _R} = {\Omega _2} = 10$ dB, and ${\Omega _E} =  - 5$ dB.}
\label{fig002}
\end{figure}
\begin{figure}[!t]
\centering
\includegraphics[width = 2.8in]{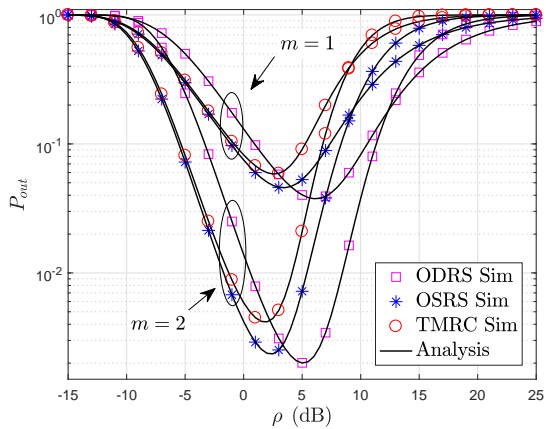}
\caption{SOP for various $m$ with ${R_1} = 0.1$, ${R_2} = 0.2$, ${m_U} = {m_E} = {m_R} = m$, ${\alpha _1} = 0.2$, ${\alpha _{\rm{J}}} = 0.5$, ${\Omega _1} = 12$ dB, ${\Omega _R} = {\Omega _2} = 10$ dB, and ${\Omega _E} =  - 5$ dB.}
\label{fig003}
\end{figure}

\section{Numerical Results and Discussions}
\label{sec:RESULTS}
In this section, we present numerical results and Monte-Carlo simulations to testify our analysis. The main adopted parameters are set to ${P_S} = {P_R} = P$, $R_1^{{\rm{th}}} = 0.2$ nat per channel use, $R_2^{{\rm{th}}} = 0.1$ nat per channel use, and ${\sigma ^2} = 1$.
In all the figures, `Sim' denotes the simulation results\footnote{We omit the results for TSRS scheme since it is exactly the same as that for OSRS.}.
One can observe that the analysis results match perfectly with simulation resutls and the secrecy performance with OSRS always outperforms that of TMRC and in some scenarios the ODRS scheme can obtain the optimal secrecy outage performance.

\begin{figure}[!t]
\centering
\includegraphics[width = 2.8in]{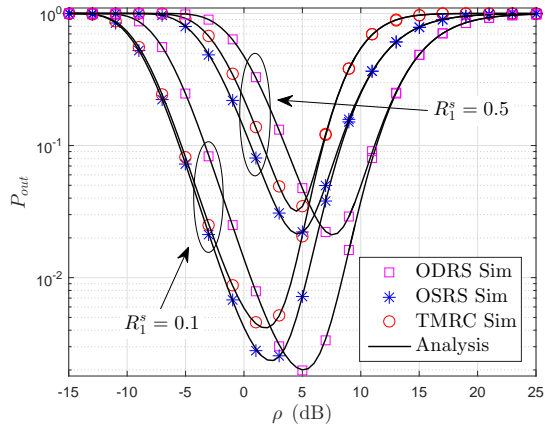}
\caption{SOP for various ${R_1^s}$ with ${R_2^s} = 0.2$, $K = 2$, ${m_U} = {m_E} = {m_R} = 2$, ${\alpha _1} = 0.2$, ${\alpha _{\rm{J}}} = 0.5$, ${\Omega _1} = 12$ dB, ${\Omega _R} = {\Omega _2} = 10$ dB, and ${\Omega _E} =  - 5$ dB.}
\label{fig004}
\end{figure}

\begin{figure}[!t]
\centering
\includegraphics[width = 2.8in]{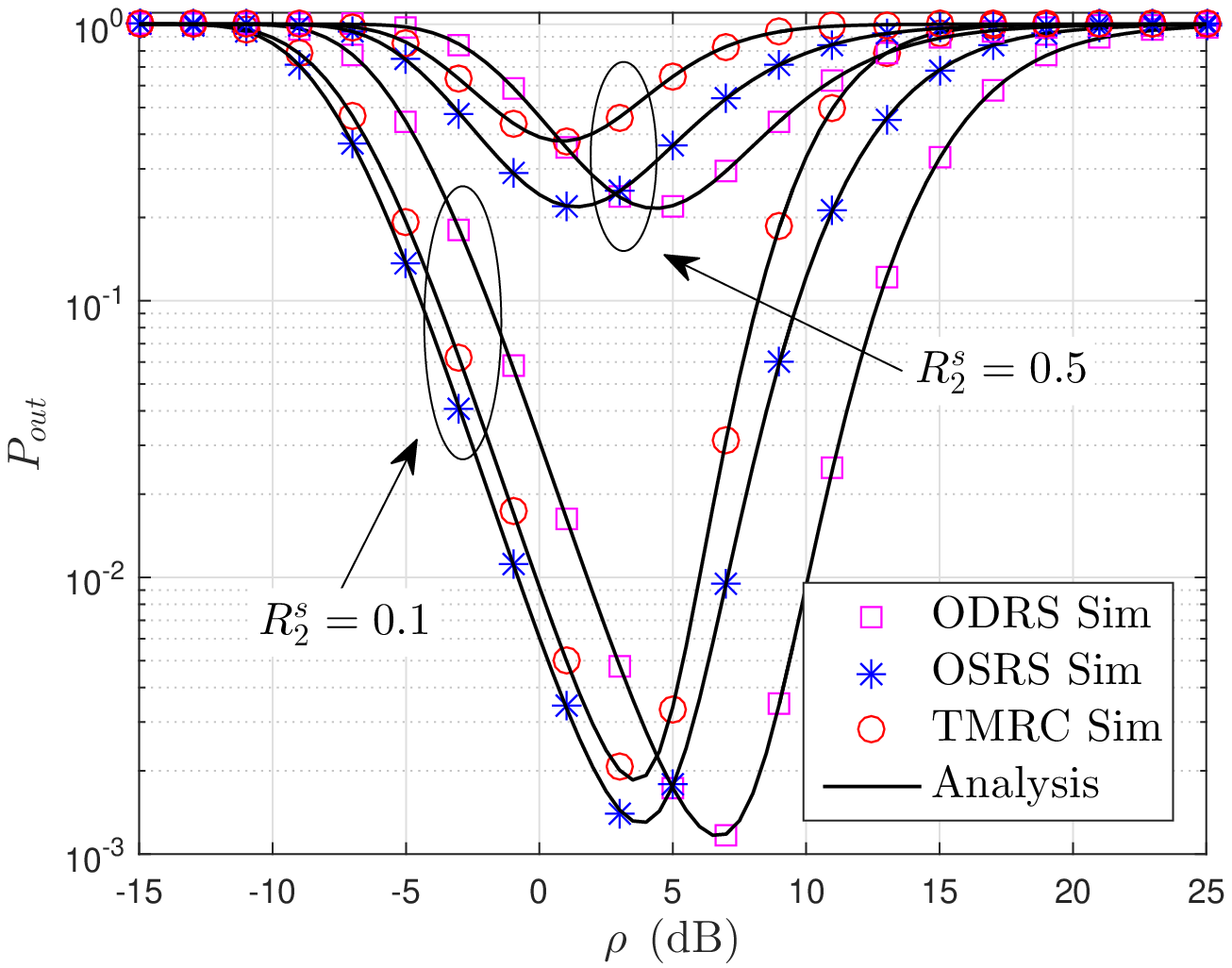}
\caption{SOP for various ${R_2^s}$ with ${R_1^s} = 0.2$, $K=2$, ${m_U} = {m_E} = {m_R} = 2$, ${\alpha _1} = 0.2$, ${\alpha _{\rm{J}}} = 0.5$, ${\Omega _1} = 12$ dB, ${\Omega _R} = {\Omega _2} = 10$ dB, and ${\Omega _E} =  - 5$ dB.}
\label{fig005}
\end{figure}

\begin{figure}[!t]
\centering
\includegraphics[width = 2.8in]{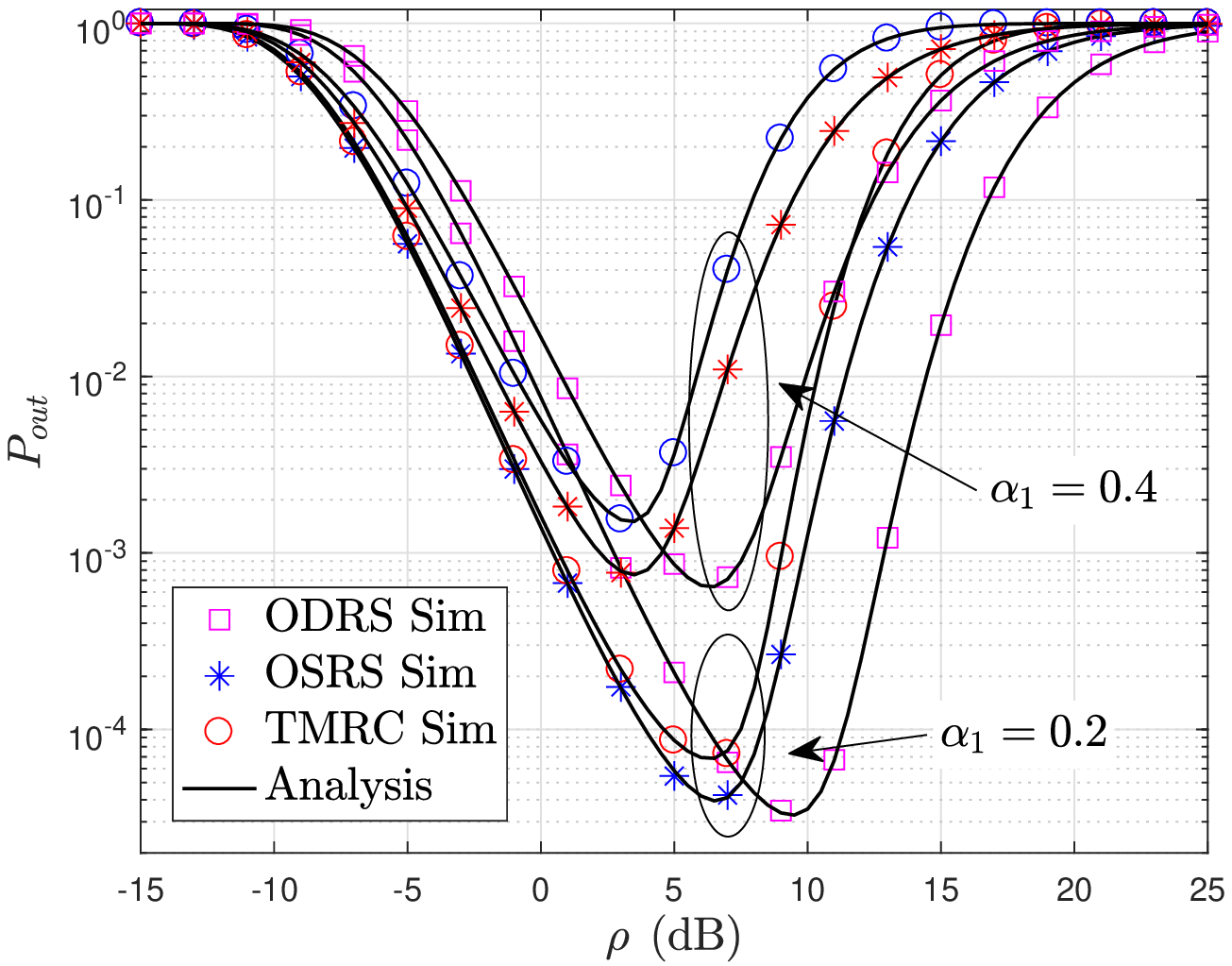}
\caption{SOP for various ${\alpha_1}$ with ${R_1} = 0.1$, ${R_2} = 0.2$, $K=2$, ${m_U} = {m_E} = {m_R} = 2$, $\alpha _{\rm{J}} = 0.5$, ${\Omega _1} = 12$ dB, ${\Omega _R} = {\Omega _2} = 10$ dB, and ${\Omega _E} =  - 10$ dB.}
\label{fig006}
\end{figure}

Figs. \ref{fig002} - \ref{fig006} present the SOP for various $\rho $ with FPA.
One can easily observe that the secrecy performance of cooperative NOMA systems is enhanced and then becomes worse by improving the transmit SNR, which is similar to the conclusions in \cite{Lei2017Access} and \cite{Lei2018TVT}. This is because the SINR of $U_2$ has a ceiling in the higher-$\rho$ region.
It is demonstrated in Fig. \ref{fig002} that the SOP is enhanced by increasing $K $ due to the improved diversity of the cooperative NOMA systems.
One can also find that the SOP depends merely on the RS scheme and is independent of $K$ in the high-$\rho$ region.
Fig. \ref{fig003} demonstrates that the SOP with larger $m$ outperforms the ones with smaller $m$ as larger $m$ signifies that the quality of wireless channel is better. However, the SOP with larger $m$ underperforms that with the smaller $m$ in the high-$\rho$ region. The reason is that the SOP of cooperative NOMA systems will benefit from the stronger fading over eavesdropping channels when the SINR of $U_2$ approaches the upper bound.

Figs. \ref{fig004} and \ref{fig005} present the effect on different ${R_1^s}$ and ${R_2^s}$, respectively.
As we know, increasing the secrecy rate threshold deteriorates the secrecy outage performance of wireless communications.
Now, from Figs. \ref{fig004} and \ref{fig005}, we can observe that ${R_1^s}$ has significant effect on the SOP of NOMA systems in low-$\rho$ region.
Different ${R_1^s}$ has significant effect on the SOP of NOMA systems in low-$\rho$ region but does not influence the SOP in high-$\rho$ region. The opposite effect is observed for ${R_2^s}$. This proves that in the low-$\rho$ region the bottleneck in secrecy performance of NOMA systems is due to $U_1$. In high-$\rho$ region, weaker user must be given more attention.
One can also observe from Fig. \ref{fig004} that the gap between two RS schemes in high-$\rho$ region is larger than those in low-$\rho$ region and the gap in low-$\rho$ region is decreased as ${R_1^s}$ decreases.
\begin{figure}[!t]
\centering
\includegraphics[width = 2.8in]{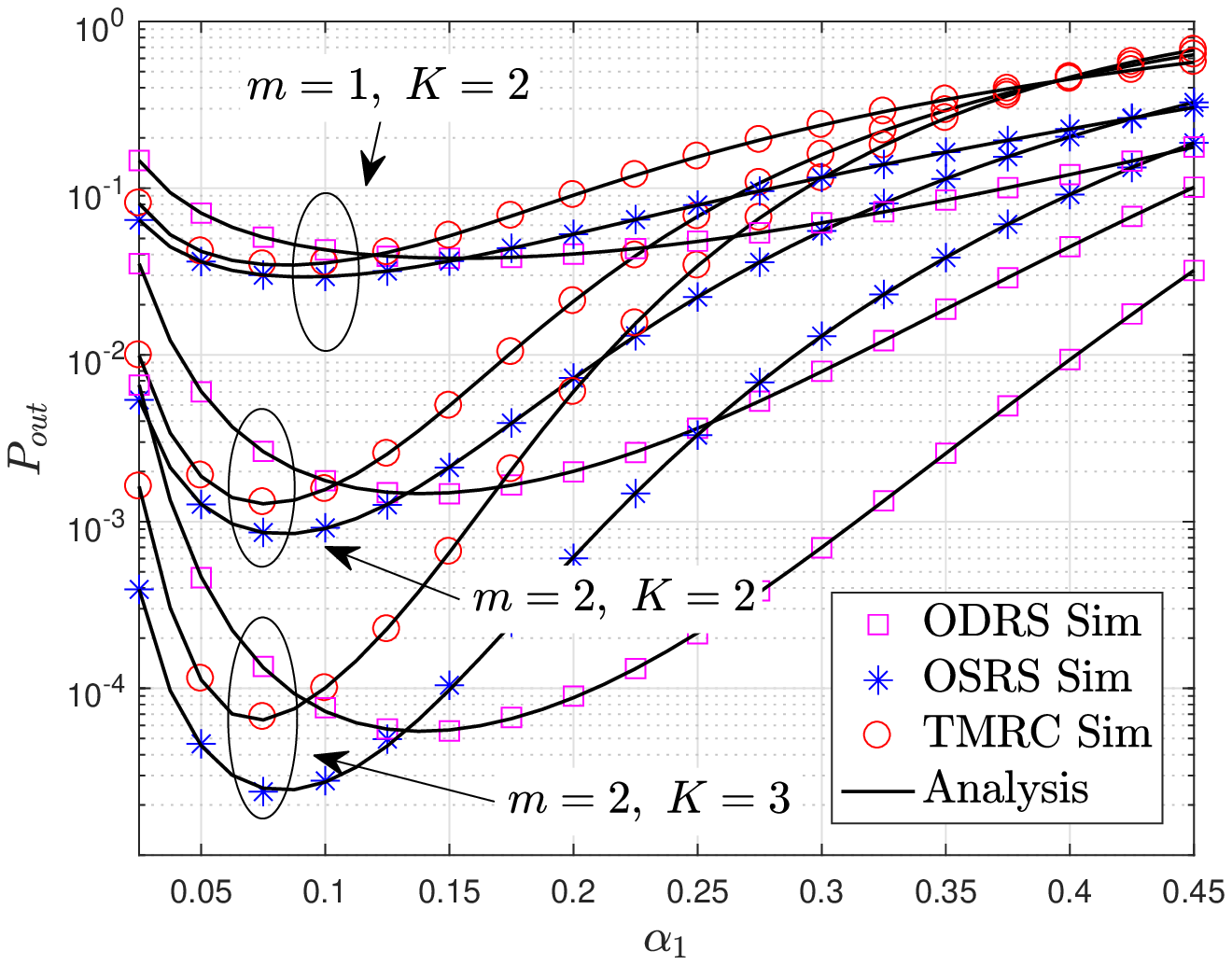}
\caption{SOP for various $\alpha _1$ and $K$ with ${R_1} = 0.1$, ${R_2} = 0.2$, ${m_U} = {m_E} = {m_R} = m$, ${\alpha _{\rm{J}}} = 0.5$, ${\Omega _1} = 12$ dB, ${\Omega _R} = {\Omega _2} = 10$ dB, and ${\Omega _E} =  - 5$ dB.}
\label{fig007}
\end{figure}
\begin{figure}[!t]
\centering
\includegraphics[width = 2.8in]{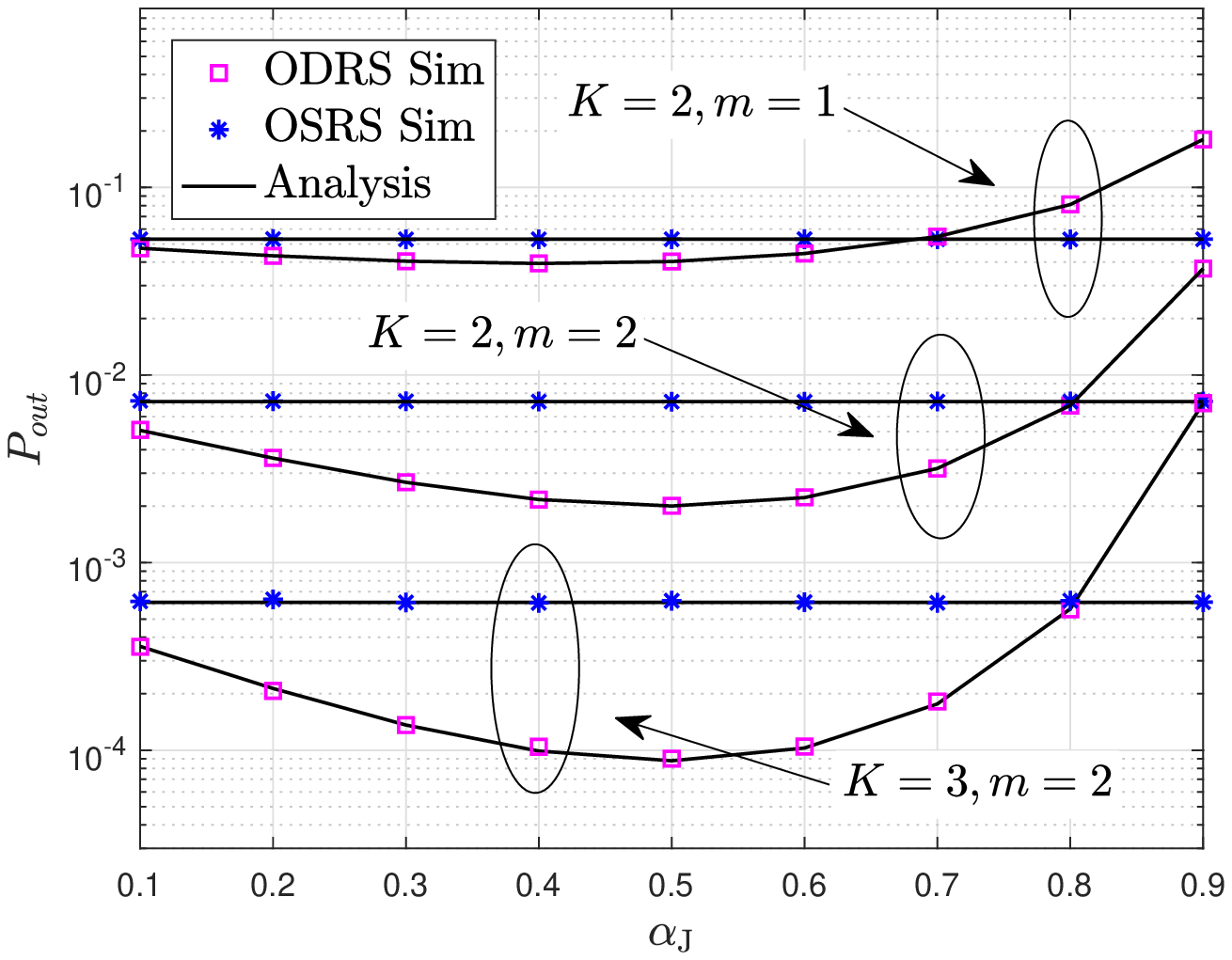}
\caption{SOP for various $\alpha _{\rm{J}}$ and $K$ and $m$ with ${R_1} = 0.1$, ${R_2} = 0.2$, ${m_U} = {m_E} = {m_R} = m $, ${\alpha _1} = 0.2$, ${\Omega _1} = 12$ dB, ${\Omega _R} = {\Omega _2} = 10$ dB, and ${\Omega _E} =  - 5$ dB.}
\label{fig008}
\end{figure}
\begin{figure}[!t]
\centering
\includegraphics[width = 2.8in]{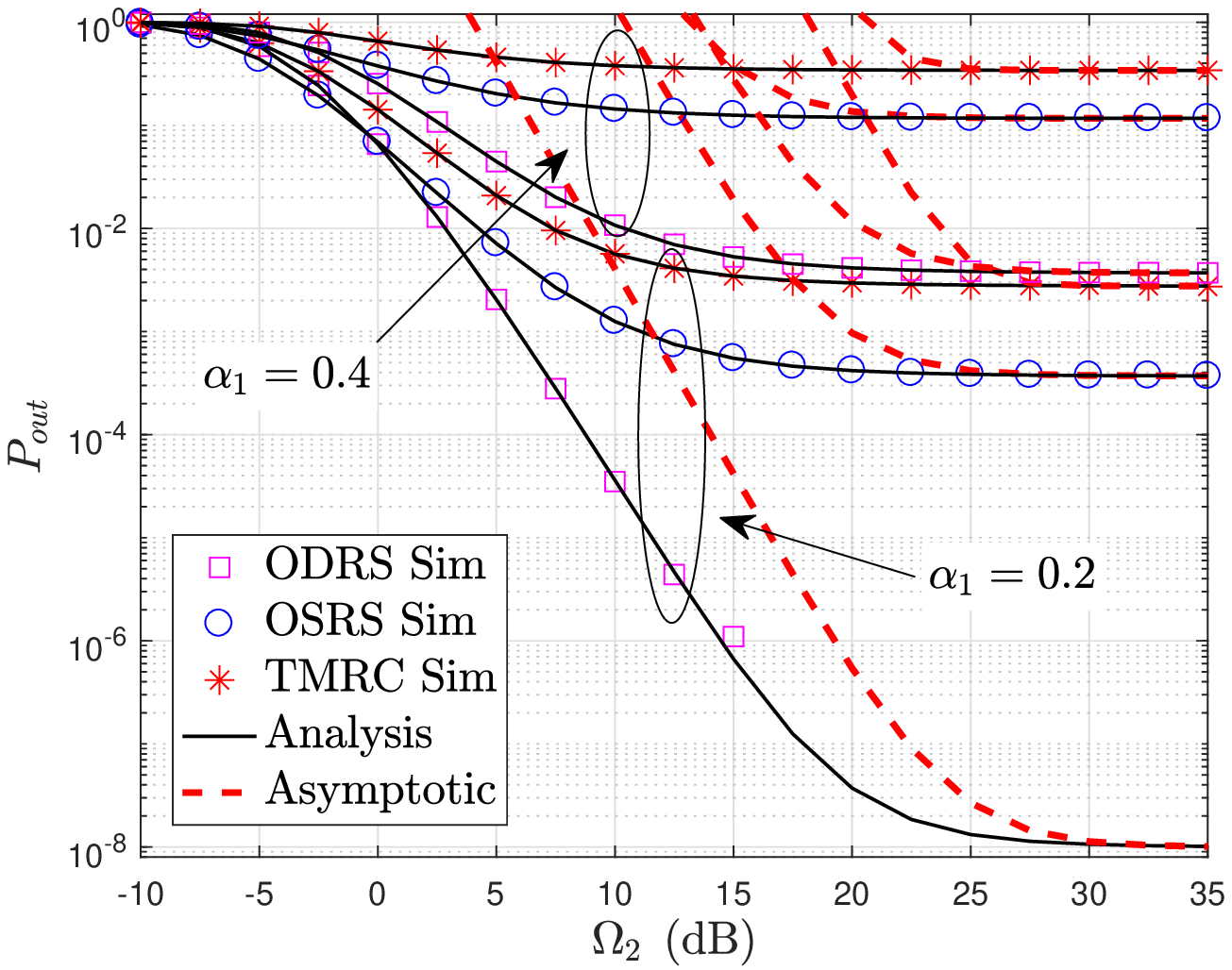}
\caption{SOP for various ${\alpha_1}$ with ${R_1} = 0.1$, ${R_2} = 0.2$, $K = 2$, $\alpha _{\rm{J}} = 0.5$, $P = 10$ dB, ${m_U} = {m_E} = {m_R} = 2$, ${\varepsilon _1} = 1.5$, ${\varepsilon _2} = 2$, and ${\Omega _E} =  - 10$ dB.}
\label{fig009}
\end{figure}

The impact of PA parameter on the SOP is investigated in Figs. \ref{fig006} - \ref{fig008}.
We can observe that the SOP decreases as ${\alpha _1}$ decreases. In other words, the secrecy performance is enhanced by enlarging the difference between allocated power to the two users, which is consistent with the results presented in Figs. \ref{fig004} and \ref{fig005}.
Fig. \ref{fig008} presents the SOP for various $\alpha _{\rm{J}}$. One can observe that the ODRS scheme enhances the secrecy outage performance of cooperative NOMA systems relative to OSRS scheme because the SNR at eavesdropper is degraded. Furthermore, the SOP is firstly improved with increasing $\alpha _{\rm{J}}$. However, it gets worse with increasing $\alpha _{\rm{J}}$ since the transmit power is reduced. Obviously, there is an optimal $\alpha _{\rm{J}}$, which can be easily obtained via one dimensional search based on the results of our work.

\begin{figure}[!t]
\centering
\includegraphics[width = 2.8in]{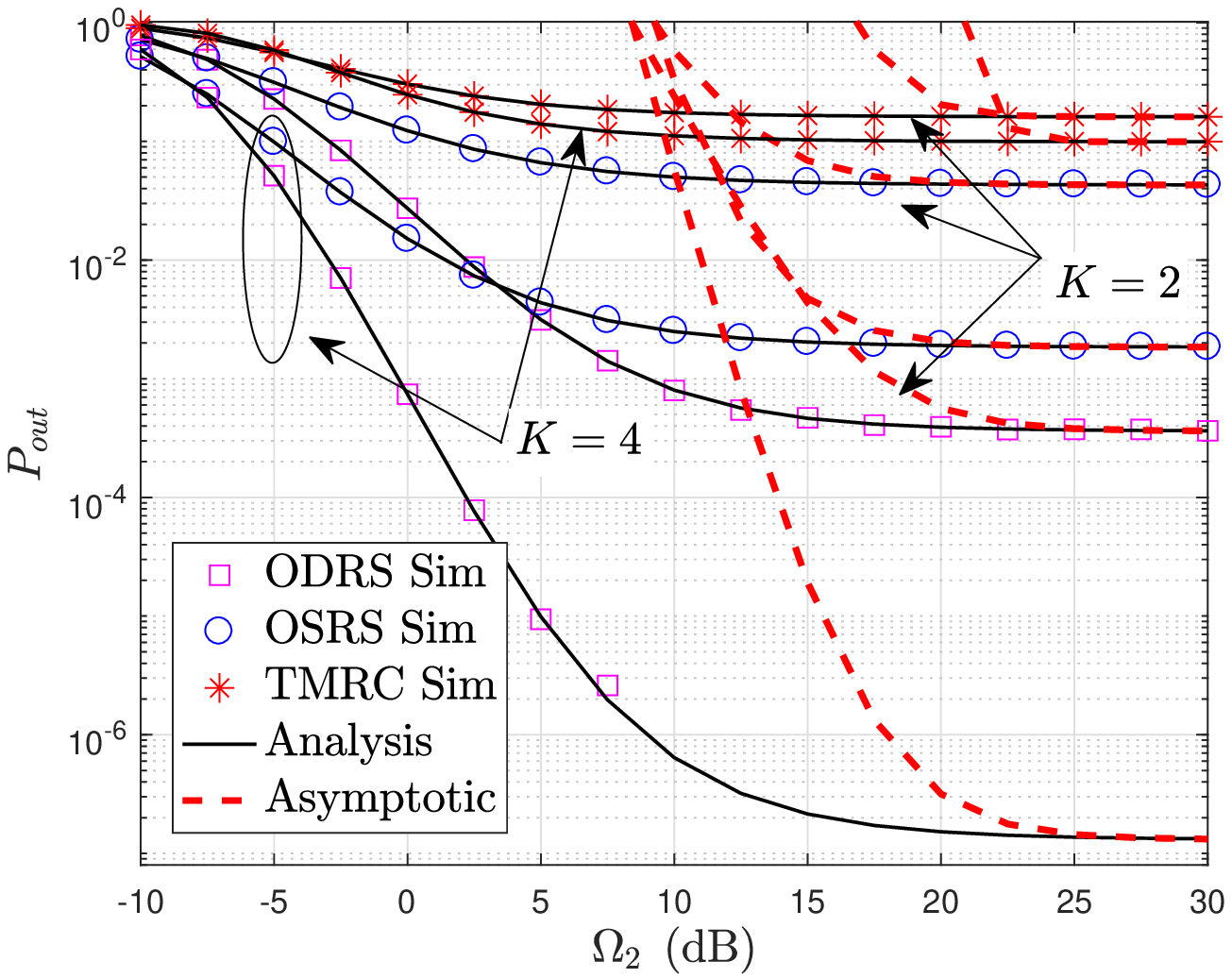}
\caption{SOP for various ${K}$ with ${R_1} = 0.1$, ${R_2} = 0.2$, ${m_U} = {m_E} = {m_R} = 2$, $P  = 15$ dB, ${\alpha_1}=0.2$, ${\alpha _{\rm{J}}} = 0.5$, ${\varepsilon _1} = 1.5$, ${\varepsilon _2} = 2$, and ${\Omega _E} =  - 12$ dB.}
\label{fig010}
\end{figure}
\begin{figure}[!t]
\centering
\includegraphics[width = 2.8in]{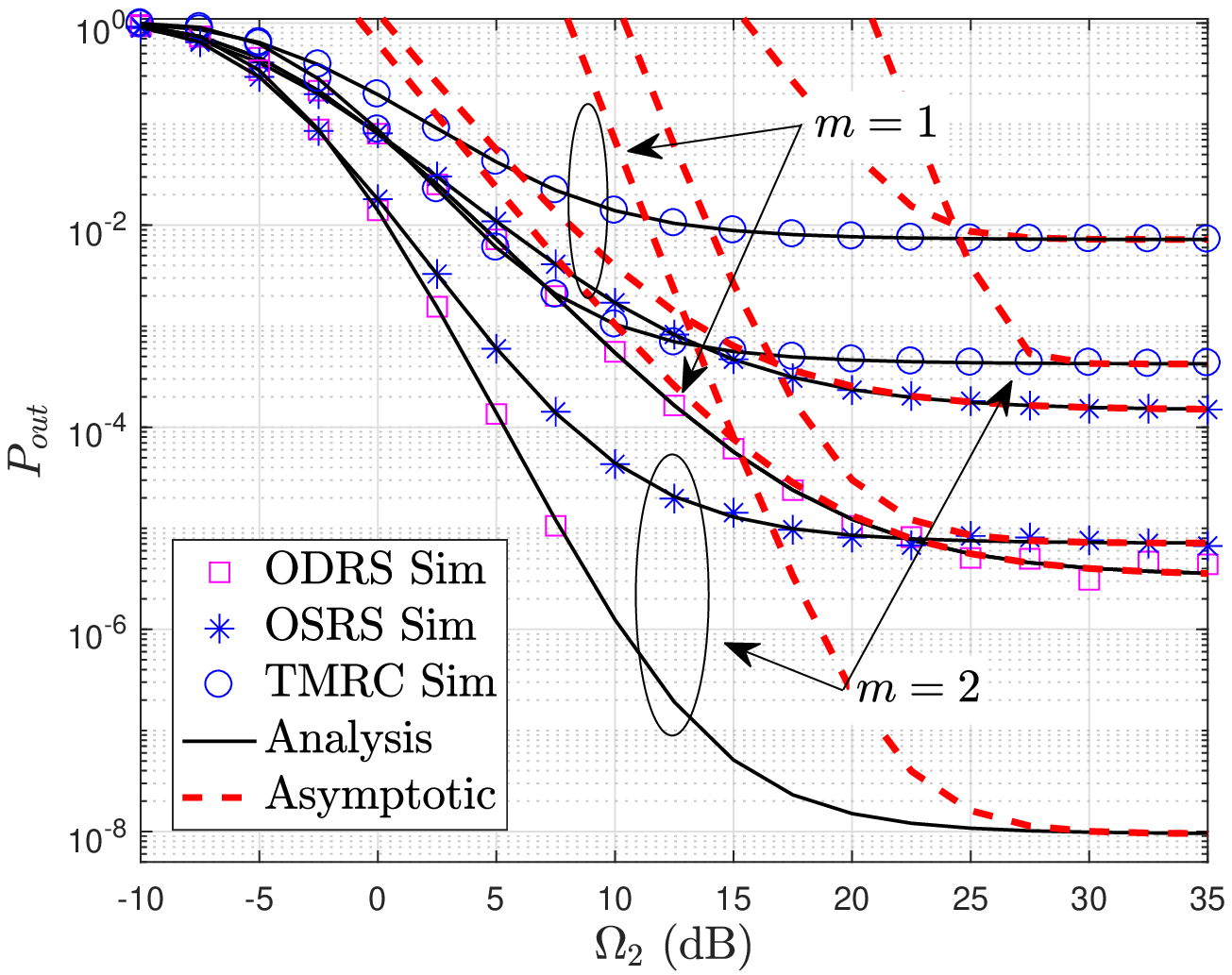}
\caption{SOP for various ${m}$ with ${R_1} = 0.1$, ${R_2} = 0.2$, ${m_U} = {m_E} = {m_R} = m$, ${\alpha _{\rm{J}}} = 0.3$, $K=3$, $P  = 10$ dB, ${\varepsilon _1} = 1.5$, ${\varepsilon _2} = 2$, and ${\Omega _E} =  - 10$ dB.}
\label{fig011}
\end{figure}

The SOP for various ${\Omega _2}$ with FPA is presented in Figs. \ref{fig009} - \ref{fig011}.
One can observe that the asymptotic SOP approaches the exact ones at high-${\Omega _2}$ region. The SDO of the cooperative NOMA systems is zero since the SOP will be a constant as ${\Omega _2}$ increases. The results in Figs. \ref{fig009} and \ref{fig010} prove that increasing $K$ and / or decreasing ${\alpha_1}$ enhances the SOP of cooperative NOMA systems. The results in Fig. \ref{fig011} demonstrate that we can obtain better secrecy outage performance under a better channel condition for both TMRC and OSRS schemes.
\begin{figure}[!t]
\centering
\includegraphics[width = 2.8in]{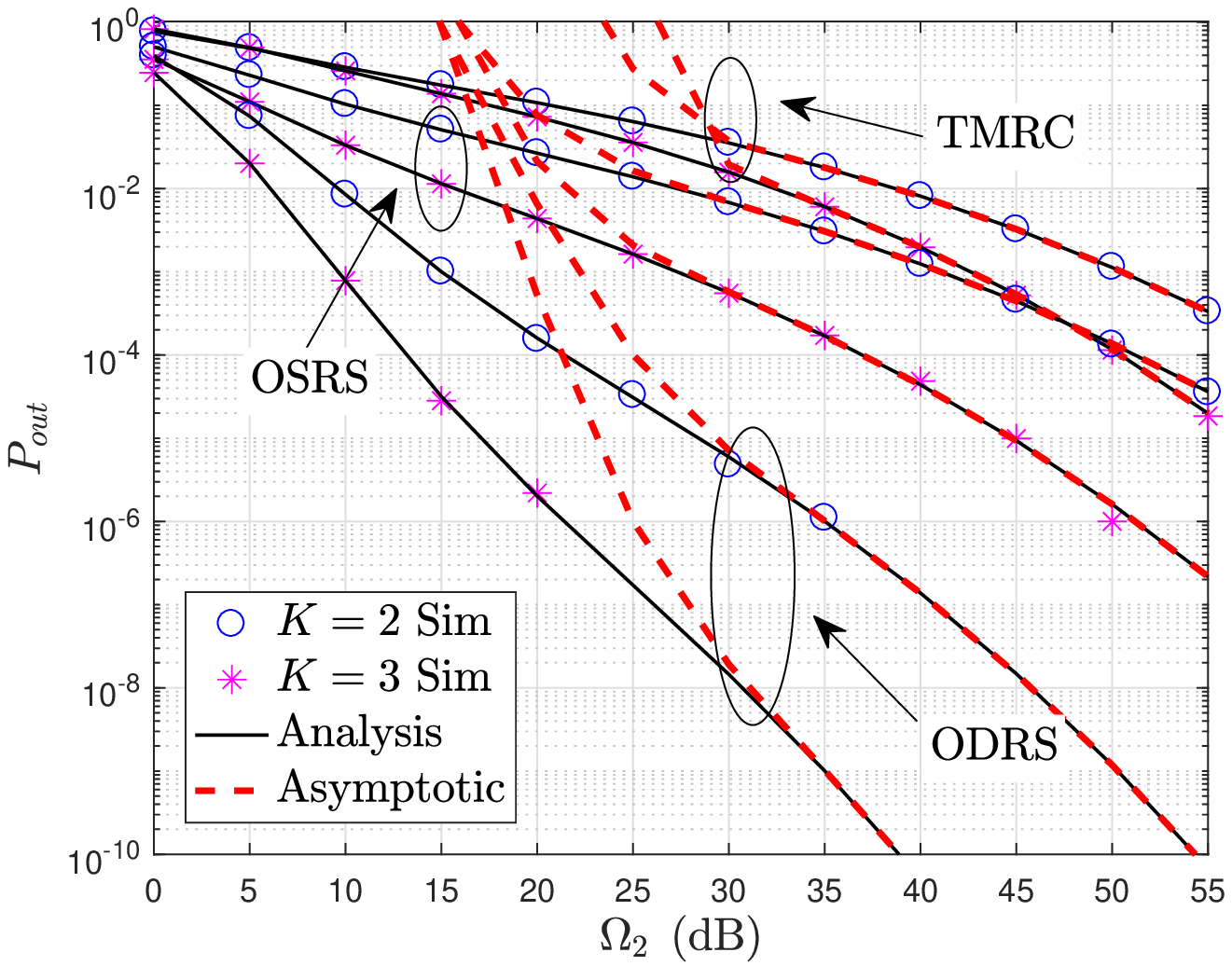}
\caption{SOP with DPA scheme for various ${K}$ with ${R_1} = 0.1$, ${R_2} = 0.2$, ${\alpha _{\rm{J}}} = 0.5$, $P = 10$ dB, ${m_U} = {m_E} = {m_R} = 2$, ${\varepsilon _1} = 1.5$, ${\varepsilon _2} = 2$, ${\varpi }=0.1$, $\mu  = 5$, and ${\Omega _E} =  - 5$ dB}.
\label{fig012}
\end{figure}
\begin{figure}[!t]
\centering
\includegraphics[width = 2.8in]{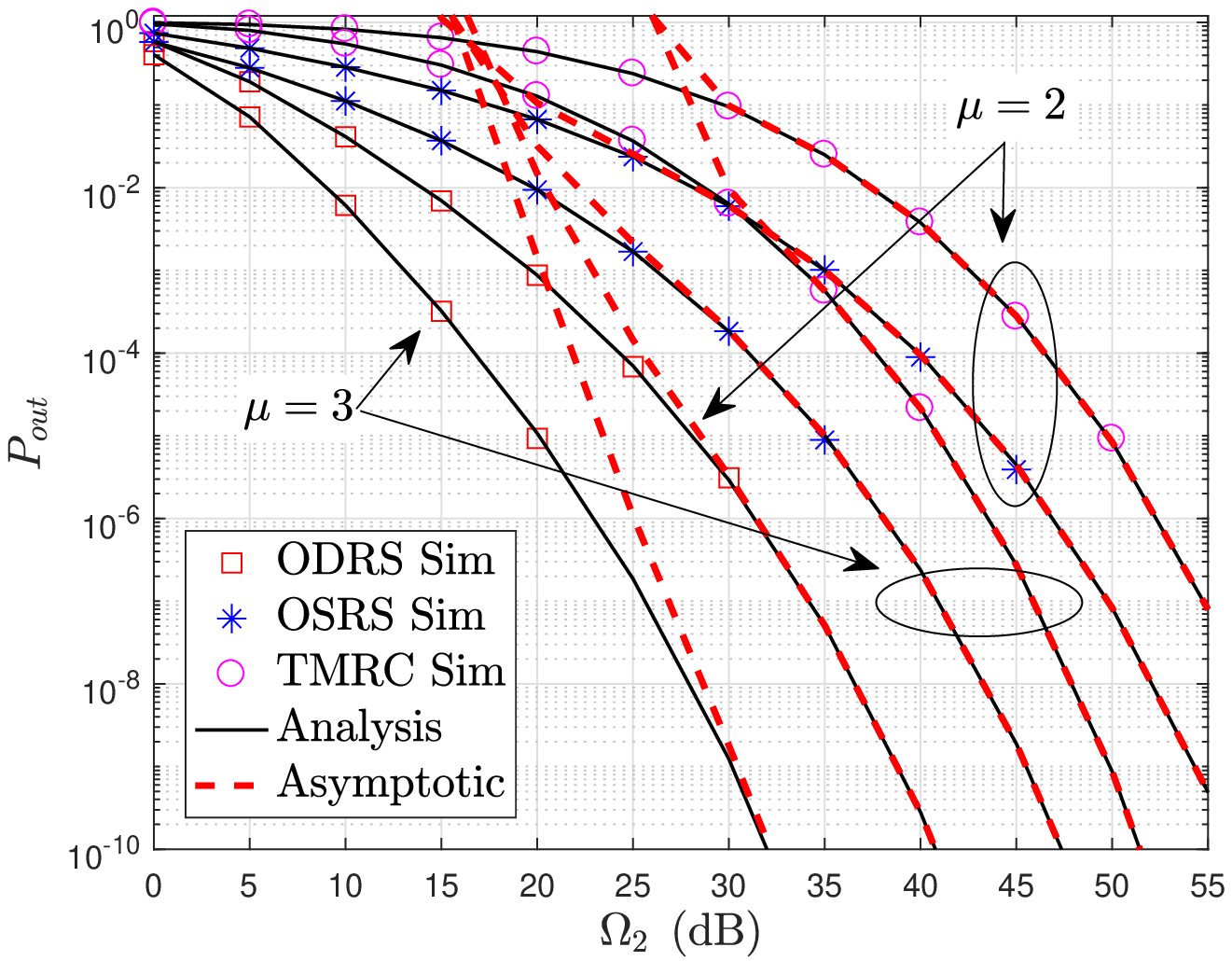}
\caption{SOP with DPA scheme for various $\mu$ with ${R_1} = 0.1$, ${R_2} = 0.2$, ${\alpha _{\rm{J}}} = 0.5$, $P = 10$ dB, ${m_U} = {m_E} = {m_R} = 2$, $K = 3$, ${\varepsilon _1} = 1.5$, ${\varepsilon _2} = 2$, $\varpi = 0.2$, and ${\Omega _E} =  - 5$ dB.}
\label{fig013}
\end{figure}

\begin{figure}[!t]
  \centering
  \subfigure[SOP with TMRC scheme]{
    \label{fig3a} 
    \includegraphics[width=1.5in]{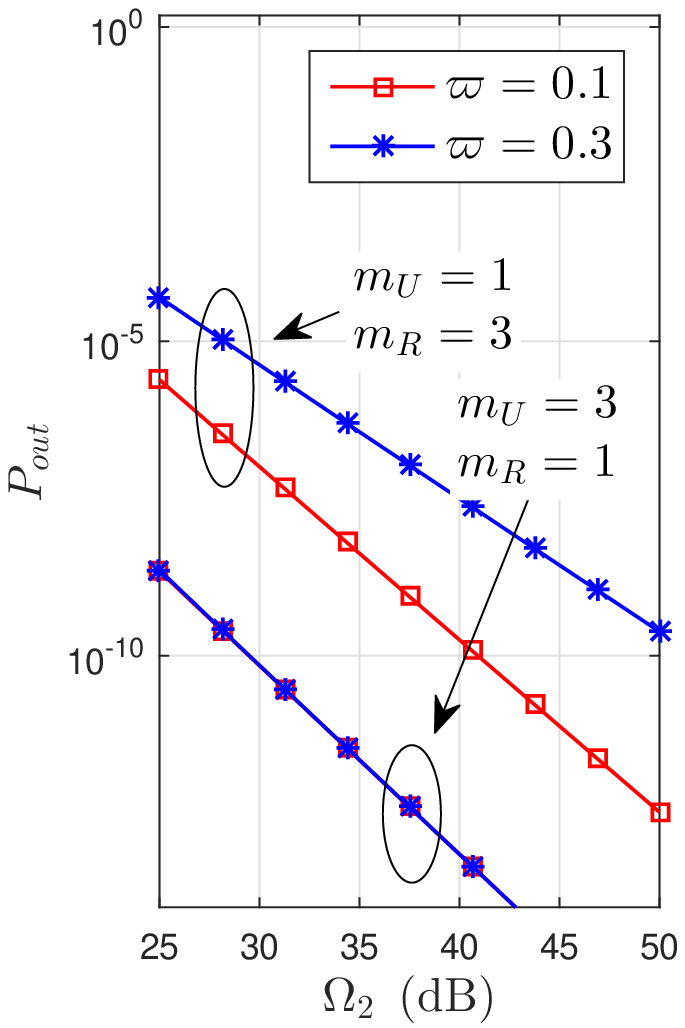}}
   \subfigure[SOP with OSRS scheme]{
    \label{fig3b} 
    \includegraphics[width=1.5in]{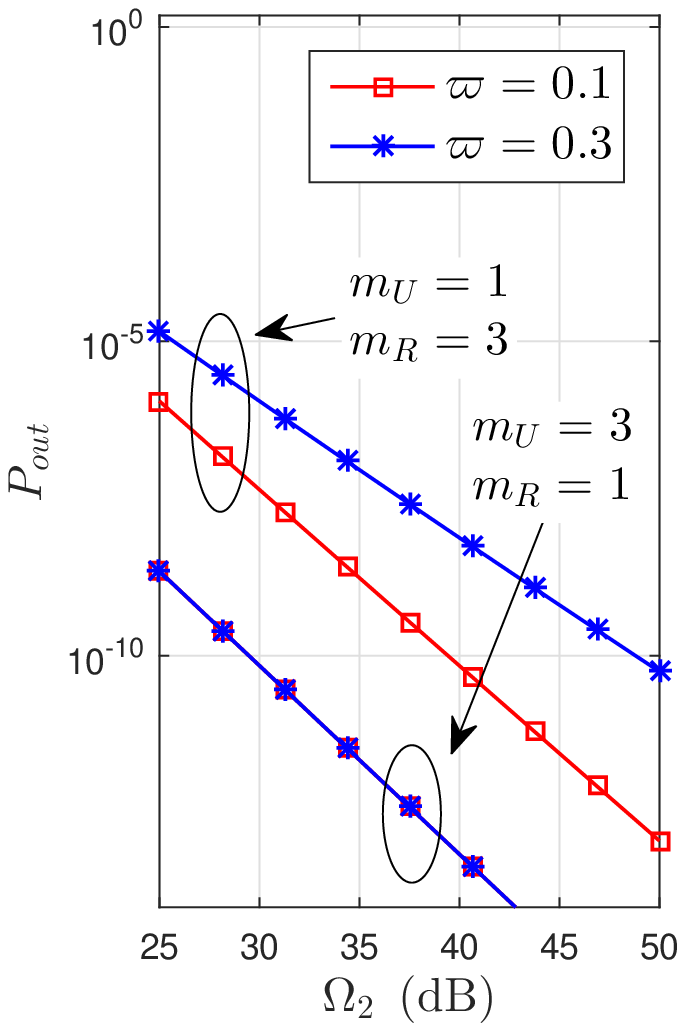}}
   \subfigure[SOP with ODRS scheme with $\varpi = 0.15$]{
    \label{fig3c} 
    \includegraphics[width=1.5in]{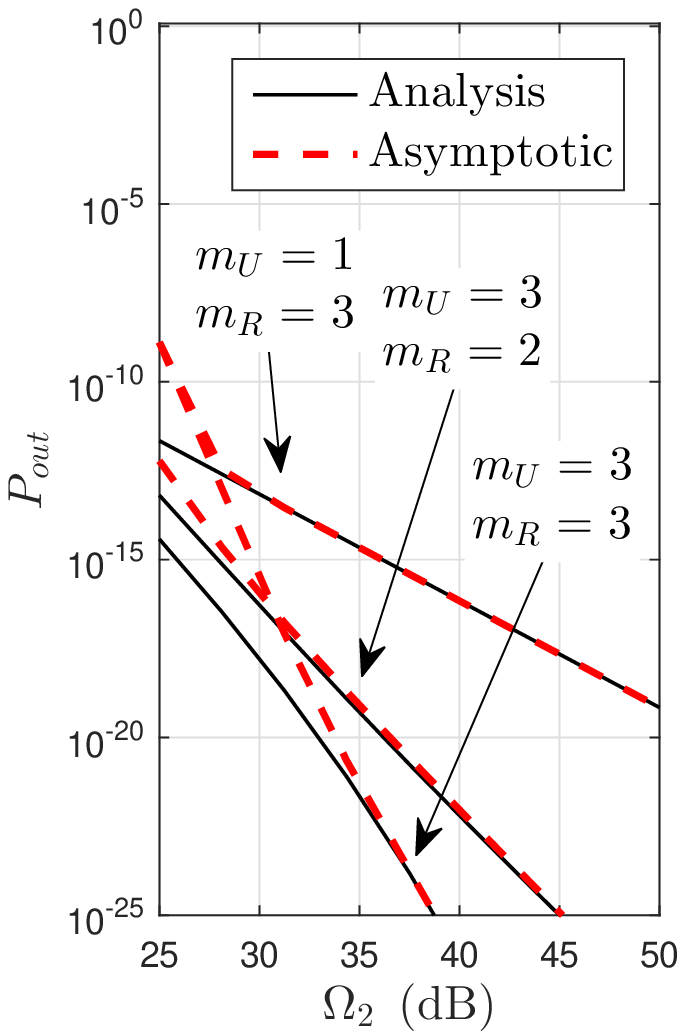}}

  \caption{SOP for DPA scheme with ${R_1} = 0.1$, ${R_2} = 0.2$, ${\alpha _{\rm{J}}} = 0.5$, $K=3$, $m_E=2$, ${\varepsilon _1} = 1.5$, ${\varepsilon _2} = 2$, ${\mu}  = 8$, and ${\Omega _E} =  - 5$ dB.}
  \label{fig014} 
\end{figure}

Figs. \ref{fig012} - \ref{fig014} present the SOP for various ${\Omega _2}$ under DPA scheme. One can observe that DPA scheme achieves a non-zero secrecy diversity order, which depends on the relay number, the fading parameters of $S - R_k$  and $S_k - U_i$. Fig. \ref{fig014} testifies that the SDO of cooperative NOMA systems equal to the minimum SDO of two hops. One can observe from Fig. \ref{fig014} (c) that a single relay is utilized to send jamming signals influences the SDO of the cooperative NOMA system in some scenarios.

\section{Conclusion}
\label{sec:Conclusion}
In this work, three RS schemes were proposed to enhance the secrecy performance of cooperative NOMA systems. We investigated the secrecy outage performance of these schemes under FPA and DPA considering the presence of dependence between the two NOMA users with different secrecy rate thresholds.
The closed-form expressions for the exact and asymptotic SOP were obtained and verified by Monte Carlo simulation results. Moreover, the SDO for all the RS schemes were derived.
The results demonstrate that relative to TMRC scheme, OSRS and ODRS schemes enhance the security of the cooperative NOMA systems. OSRS scheme achieves the same SDO as TMRC scheme. The jamming signal influences the SDO of ODRS scheme. There are optimal PA coefficients between jamming and signal transmissions, which can be easily obtained via one dimensional search based on the results of our work.

\begin{appendices}
\section{}
The PDF of ${H_E} = \mathop {\max }\limits_{k \in \bar \Phi } \left\{ {G_E^k} \right\}$ is obtained as
\setcounter{equation}{52}
\begin{equation}
{f_{{H_E}}}\left( z \right) = \left( {K - n} \right){\left( {{F_{{G_E}}}\left( z \right)} \right)^{K - n - 1}}{f_{{G_E}}}\left( z \right),
\label{auhn}
\end{equation}
where ${{F_{{G_E}}}\left( z \right)}$ and ${{f_{{G_E}}}\left( z \right)}$ refer to the CDF and PDF, respectively, of ${G_E^k}$, which is easily obtained by utilizing (\ref{EPDFMR}).
Based on multi-nomial theorem, we obtain the pdf of ${H_E}$ as
\begin{equation}
{f_{{H_E}}}\left( z \right) = \frac{{(K - n){\lambda _E^{{m_E}}}}}{{\Gamma \left( {{m_E}} \right)}}\sum\limits_{S_E} {A{z^{B + {m_E} - 1}}{e^{ - C{\lambda _E}z}}},
\end{equation}
where ${S_E} = \left\{ {\left( {{n_1}, \cdots ,{n_{{m_E} + 1}}} \right) \in {\Bbb{N}}\left. {\left| {\sum\limits_{p = 1}^{{m_E} + 1} {{n_p}}  = K - n - 1} \right.} \right|} \right\}$,
$A = \left( {\frac{{\left( {K - n - 1} \right)!}}{{\prod\limits_{q = 1}^{{m_E} + 1} {\left( {{n_q}} \right)!} }}} \right)\prod\limits_{p = 2}^{{m_E} + 1} {{{\left( { - \frac{{\lambda _E^{p - 2}}}{{\left( {p - 2} \right)!}}} \right)}^{{n_p}}}} $,
$B = \sum\limits_{p = 2}^{{m_E} + 1} {{n_p}\left( {p - 2} \right)} $,
and
$C = 1 + \sum\limits_{p = 2}^{{m_E} + 1} {{n_p}}$.

\section{}
Based on (\ref{pdfHE}) and utilizing (1.111) and (3.351.3) of \cite{Gradshteyn2014}, the CDF of $Y$ can be obtained as
\begin{equation}
\begin{aligned}
{F_Y}\left( y \right) &= \Pr \left\{ {\frac{{G_E^k}}{{1 + {\rho _4}{H_E}}} < y} \right\}\\
& = \Pr \left\{ {G_E^k < y + {\rho _4}{H_E}y} \right\}\\
&  = \int_0^\infty  {{F_{G_E^k}}\left( {y + {\rho _4}xy} \right){f_{{H_E}}}\left( x \right)dx} \\
& = 1 - \varphi \sum\limits_{k = 0}^{{m_E} - 1} {\sum\limits_{{S_E}} {\frac{{A\lambda _E^k{e^{ - {\lambda _E}y}}{y^k}}}{{k!}}\int_0^\infty  {{x^{B + {m_E} - 1}}{{\left( {{\rho _4}x + 1} \right)}^k}{e^{ - {\lambda _E}\left( {C + {\rho _4}y} \right)x}}dx} } } \\
& = 1 - \varphi \sum\limits_{k = 0}^{{m_E} - 1} {\sum\limits_{S_E} {\sum\limits_{j = 0}^k {\frac{{C_k^jA\rho _4^j\lambda _E^k{e^{ - {\lambda _E}y}}{y^k}}}{{k!}}\int_0^\infty  {{x^{B + j + {m_E} - 1}}{e^{ - {\lambda _E}\left( {C + {\rho _4}y} \right)x}}dx} } } } \\
& = 1 - \varphi \sum\limits_{k = 0}^{{m_E} - 1} {\sum\limits_{S_E} {\sum\limits_{j = 0}^k {\frac{{\delta {e^{ - {\lambda _E}y}}{y^k}}}{{{{\left( {C + {\rho _4}y} \right)}^\varsigma }}}} } },
\label{cdfY}
\end{aligned}
\end{equation}
where $\varphi  = \frac{{\left( {K - n} \right)\lambda _E^{{m_E}}}}{{\Gamma \left( {{m_E}} \right)}}$,
$\varsigma = B + {m_E} + j$,
and
$\delta  = \frac{{C_k^jA\rho _4^j\lambda _E^{k - \varsigma }\left( {\varsigma  - 1} \right)!}}{{k!}}$.

Using integration by parts, and applying some algebra, we derive the PDF of $Y$ as (\ref{pdfY}).

\end{appendices}


\begin{thebibliography}{1}

\bibitem{DaiL2015Mag}
L. Dai, B. Wang, Y. Yuan, S. Han, C.-L. I, and Z. Wang, ``Non-orthogonal multiple access for 5G: solutions, challenges, opportunities, and future research trends," \emph{IEEE Commun. Mag.}, vol. 53, no. 9, pp. 74-81, Sept. 2015.

\bibitem{WangY2016CC}
Y. Wang, B. Ren, S. Sun, S. Kang and X. Yue, ``Analysis of non-orthogonal multiple access for 5G," \emph{China Communications}, vol. 13, no. Supplement2, pp. 52-66, N/A 2016.

\bibitem{DingZ2017Mag}
Z. Ding, Y. Liu, J. Choi, Q. Sun, M. Elkashlan, I. C. L, and H. V. Poor, ``Application of non-orthogonal multiple access in LTE and 5G networks," \emph{IEEE Commun. Mag.}, vol. 55, no. 2, pp. 185-191, Feb. 2017.

\bibitem{IslamSMR2017Survey}
S. M. R. Islam, N. Avazov, O. A. Dobre, and K. S. Kwak, ``Power-domain non-orthogonal multiple access (NOMA) in 5G systems: Potentials and challenges," \emph{IEEE Commun. Surveys Tuts.}, vol. 19, no. 2, pp. 721 - 742, 2nd Quart. 2017.

\bibitem{Liu2017Proc}
Y. Liu, Z. Qin, M. Elkashlan, Z. Ding, A. Nallanathan, and L. Hanzo, ``Nonorthogonal multiple access for 5G and beyond," \emph{Proc. IEEE,} vol. 105, no. 12, pp. 2347-2381, Dec. 2017.

\bibitem{Ding2017JSAC}
Z. Ding, X. Lei, G. K. Karagiannidis, R. Schober, J. H. Yuan, and V. K. Bhargava, ``A survey on non-orthogonal multiple access for 5G networks: Research challenges and future trends," \emph{IEEE J. Sel. Areas Commun.,} vol. 35, no. 10, pp. 2181-2195, Oct. 2017.

\bibitem{Liu2018WC}
Y. Liu, H. Xing, C. Pan, A. Nallanathan, M. Elkashlan, and L. Hanzo, ``Multiple antenna assisted non-orthogonal multiple access," \emph{IEEE Wireless Commun.}, vol. 25, no. 2, pp. 17 - 23, Apr. 2018.

\bibitem{WanD2018WC}
D. Wan, M. Wen, F. Ji, H. Yu, and F. Chen, ``Non-orthogonal multiple access for cooperative communications: Challenges, opportunities, and trends," \emph{IEEE Wireless Commun.}, vol. 25, no. 2, pp. 109-117, Apr. 2018.

\bibitem{ZhouF2018WC}
F. Zhou, Y. Wu, Y.-C. Liang, Z. Li, Y. Wang, and K.-K. Wong, ``State of the art, taxonomy, and open issues on cognitive radio networks with NOMA," \emph{IEEE Wireless Commun.}, vol. 25, no. 2, pp. 100-108, Apr. 2018.


\bibitem{Ding2015CL}
Z. Ding, M. Peng, and H. V. Poor, ``Cooperative non-orthogonal multiple access in 5G systems," \emph{IEEE Commun. Lett.,} vol. 19, no. 8, pp. 1462-1465, Aug. 2015.

\bibitem{LiuY2016JSACSWIPT}
Y. Liu, Z. Ding, M. Elkashlan, and H. V. Poor, ``Cooperative non-orthogonal multiple access with simultaneous wireless information and power transfer," \emph{IEEE J. Sel. Areas Commun.}, vol. 34, no. 4, pp. 938-953, Apr. 2016.


\bibitem{MenJ2015CL}
J. Men and J. Ge, ``Non-orthogonal multiple access for multiple-antenna relaying networks," \emph{IEEE Commun. Lett.}, vol. 19, no. 10, pp. 1686-1689, Oct. 2015.

\bibitem{MenJ2017TVT}
J. Men, J. Ge, and C. Zhang, ``Performance analysis of non-orthogonal multiple access for relaying networks over Nakagami-$m$ fading channels," \emph{IEEE Trans. Veh. Technol.}, vol. 66, no. 2, pp. 1200-1208, Feb. 2017.


\bibitem{XiaoY2018CL}
Y. Xiao, L. Hao, Z. Ma, Z. Ding, Z. Zhang, and P. Fan, ``Forwarding strategy selection in dual-hop NOMA relaying systems," \emph{IEEE Commun. Lett.}, vol. 22, no. 8, pp. 1644-1647, Aug. 2018.


\bibitem{Duan2017Access}
W. Duan, M. Wen, Z. Xiong, and M. H. Lee, ``Two-stage power allocation for dual-hop relaying systems with non-orthogonal multiple access," \emph{IEEE Access}, vol. 5, pp. 2254-2261, Mar. 2017.

\bibitem{ZhongC2016CL}
C. Zhong and Z. Zhang, ``Non-Orthogonal multiple access with cooperative full-duplex relaying," \emph{IEEE Commun. Lett.}, vol. 20, no. 12, pp. 2478-2481, Dec. 2016.

\bibitem{ZhangZ2016TVT}
Z. Zhang, Z. Ma, M. Xiao, Z. Ding, and P. Fan, ``Full-duplex device-to-device aided cooperative non-orthogonal multiple access," \emph{IEEE Trans. Veh. Technol.}, vol. 66, no. 5, pp. 4467 - 4471, May 2016.

\bibitem{LiuG2018TWC}
G. Liu, X. Chen, Z. Ding, Z. Ma, and F. R. Yu, ``Hybrid half-duplex/full-duplex cooperative non-orthogonal multiple
access with transmit power adaptation," \emph{IEEE Trans. Wireless Commun.}, vol. 17, no. 1, pp. 506-519, Jan. 2018.

\bibitem{ChuTMC2018CL}
T. M. C. Chu and H.-J. Zepernick, ``Performance of a non-orthogonal multiple access system with full-duplex relaying," \emph{IEEE Commun. Lett.}, vol. 22, no. 10, pp. 2084-2087, Oct. 2018.

\bibitem{KaderMF2018TVT}
M. F. Kader, S. Y. Shin, and V. C. M. Leung, ``Full-duplex non-orthogonal multiple access in cooperative relay sharing for 5G systems," \emph{IEEE Trans. Veh. Technol.}, vol. 67, no. 6, pp. 5831-5840, Jul. 2018.

\bibitem{ZouY2013JSAC}
Y. Zou, X. Wang, and W. Shen, ``Optimal relay selection for physical-layer security in cooperative wireless networks," \emph{IEEE J. Sel. Areas Commun.}, vol. 31, no. 10, pp. 2099-2111, Oct. 2013.

\bibitem{LeiH2017TCCN}
H. Lei, H. Zhang, I. S. Ansari, Z. Ren, G. Pan, K. A. Qaraqe, and M.-S. Alouini, ``On secrecy outage of relay selection in underlay cognitive radio networks over Nakagami-m fading channel," \emph{IEEE Trans. Cogn. Commun. Netw.}, vol. 3, no. 4, pp. 614 - 627, Dec. 2017.

\bibitem{Ding2016WCL}
Z. Ding, H. Dai, and H. V. Poor, ``Relay selection for cooperative NOMA," \emph{IEEE Wireless Commun. Lett.,} vol. 5, no. 4, pp. 416-419, Aug. 2016.

\bibitem{ZhaoJ2018access}
J. Zhao, Z. Ding, P. Fan, Z. Yang, and G. K. Karagiannidis, ``Dual relay selection for cooperative NOMA with distributed space time coding," \emph{IEEE Access}, vol. 6, pp. 20440-20450, Apr. 2018.

\bibitem{YangZ2017TVT}
Z. Yang, Z. Ding, Y. Wu, and P. Fan, ``Novel relay selection strategies for cooperative NOMA," \emph{IEEE Trans. Veh. Technol.,} vol. 66, no. 11, pp. 10114-10123, Nov. 2017.

\bibitem{XuP2018TVT}
P. Xu, Z. Yang, Z. Ding, and Z. Zhang, ``Optimal relay selection schemes for cooperative NOMA,'' \emph{IEEE Trans. Veh. Technol.}, vol. 67, no. 8, pp. 7851-7855, Aug. 2018.

\bibitem{YueX2018TCOM}
X. Yue, Y. Liu, S. Kang, A. Nallanathan, and Z. Ding, ``Spatially random relay selection for full/half-duplex cooperative NOMA networks," \emph{IEEE Trans. Commun.}, vol. 66, no. 8, pp. 3294 - 3308, Aug. 2018.

\bibitem{ChenJ2018Access}
J. Chen, L. Yang, and M.-S. Alouini, ``Performance analysis of cooperative NOMA schemes in spatially random relaying networks," \emph{IEEE Access}, vol. 6, pp. 33159-33168, Jul. 2018.

\bibitem{Zou2016Proc}
Y. Zou, J. Zhu, X. Wang, and L. Hanzo, ``A survey on wireless security: Technical challenges, recent advances, and future trends," \emph{Proc. IEEE}, vol. 104, no. 9, pp. 1727-1765, May 2016.

\bibitem{WangHM2016Book}
H.-M. Wang and T.-X. Zheng, \emph{Physical Layer Security in Random Cellular Networks}. Singapore: Springer, 2016.

\bibitem{ZhangY2016CL}
Y. Zhang, H. Wang, Q. Yang, and Z. Ding, ``Secrecy sum rate maximization in non-orthogonal multiple access," \emph{IEEE Commun. Lett.,} vol. 20, no. 5, pp. 930-933, May. 2016.

\bibitem{LiuY2017TWC}
Y. Liu, Z. Qin, M. Elkashlan, Y. Gao, and L. Hanzo, ``Enhancing the physical layer security of non-orthogonal multiple access in large-scale networks,''\emph{IEEE Trans. Wireless Commun.}, vol. 16, no. 3, pp. 1656-1672, Mar. 2017.

\bibitem{Lei2017Access}
H. Lei, J. Zhang, K. Park, P. Xu, I. S. Ansari, G. Pan, B. Alomair, and M.-S. Alouini, ``On secure NOMA systems with transmit antenna selection schemes,''\emph{IEEE Access,} vol. 5, pp. 17450-17464, Sep. 2017.

\bibitem{Lei2018TVT}
H. Lei, J. Zhang, K.-H. Park, P. Xu, Z. Zhang, G. Pan, and M.-S. Alouini, ``Secrecy outage of max-min TAS scheme in MIMO-NOMA systems,''\emph{IEEE Trans. Veh. Technol.}, vol. 67, no. 8, pp. 6981-6990, Aug. 2018.

\bibitem{FengY2018arXiv}
Y. Feng, S. Yan, and Z. Yang. ``Secure transmission to the stronger user with optimal power allocation in NOMA," arXiv:1807.00935, Jul. 2018, [Online]: https://arxiv.org/abs/1807.00935

\bibitem{LvL2018TVT}
L. Lv, Z. Ding, Q. Ni, and J. Chen, ``Secure MISO-NOMA transmission with artificial noise," \emph{IEEE Trans. Veh. Technol.}, vol. 67, no. 7, pp. 6700-6705, Jul. 2018.


\bibitem{ZhouF2018JSAC}
F. Zhou, Z. Chu, H. Sun, R. Q. Hu, and L. Hanzo, ``Artificial noise aided secure cognitive beamforming for cooperative MISO-NOMA using SWIPT," \emph{IEEE J. Sel. Areas Commun.}, vol. 36, no. 4, pp. 918-931, Apr. 2018.

\bibitem{ZhuZ2018WCL}
Z. Zhu, Z. Chu, F. Zhou, H. Niu, Z. Wang, and I. Lee, ``Secure beamforming designs for secrecy MIMO SWIPT systems," \emph{IEEE Wireless Commun. Lett.}, vol. 7, no. 3, pp. 424-427, Jun. 2018.

\bibitem{JiangK2018Access}
K. Jiang, T. Jing, Y. Huo, F. Zhang, and Z. Li, ``SIC-based secrecy performance in uplink NOMA multi-eavesdropper wiretap channels," \emph{IEEE Access}, vol. 6, pp. 19664-19680, Apr. 2018.

\bibitem{ZhangH2018JSAC}
H. Zhang, N. Yang, K. Long, M. Pan, G. K. Karagiannidis, and V. C. M. Leung, ``Secure communications in NOMA system: Subcarrier assignment and power allocation," \emph{IEEE J. Sel. Areas Commun.}, vol. 36, no. 7, pp. 1441-1452, Jul. 2018.


\bibitem{Chen2018TVT}
J. Chen, L. Yang, and M. Alouini, ``Physical layer security for cooperative NOMA systems,''\emph{IEEE Trans. Veh. Technol.}, vol. 67, no. 5, pp. 4645-4649, May. 2018.

\bibitem{JiangW2018TVT}
W. Jiang, Y. Gong, Q. Xiao, and Y. Liao, ``Rate maximization for untrusted relay networks with nonorthogonal cooperative transmission protocols," \emph{IEEE Trans. Veh. Technol.}, vol. 67, no. 7, pp. 6325-6339, Jul. 2018.


\bibitem{Arafa2018arXiv}
A. Arafa, W. Shin, M. Vaezi, and H. V. Poor. ``Secure relaying in non-orthogonal multiple access: Trusted and untrusted scenarios," arXiv:1808.07864, Aug. 2018, [Online]: https://arxiv.org/abs/1808.07864

\bibitem{FengY2017GC}
Y. Feng, Z. Yang, and S. Yan, ``Non-orthogonal multiple access and artificial-noise aided secure transmission in FD relay networks," in \emph{Proc. 2017 IEEE Globecom Workshops (GC Wkshps)}, Singapore, Dec. 2017, pp. 1-6.

\bibitem{LiuC2018ICC}
C. Liu, L. Zhang, M. Xiao, Z. Chen, and S. Li, ``Secrecy performance analysis in downlink NOMA systems with cooperative full-duplex relaying," in \emph{Proc. 2018 IEEE International Conference on Communications Workshops (ICC Workshops)}, Kansas City, MO, USA, Jul. 2018, pp. 1-6.

\bibitem{LiB2018TCOM}
B. Li, X. Qi, K. Huang, Z. Fei, F. Zhou, and R. Q. Hu, ``Security-reliability tradeoff analysis for cooperative NOMA in cognitive radio networks," \emph{IEEE Trans. Commun.}, doi: 10.1109/tcomm.2018.2873690, Oct. 2018.


\bibitem{Gradshteyn2014}
I. S. Gradshteyn and I. M. Ryzhik, \emph{Table of Integrals, Series, and Products}: Academic press, 2014.

\bibitem{Lei2017TVTTAS}
H. Lei, C. Gao, I. S. Ansari, Y. Guo, Y. Zou, G. Pan, and K. Qaraqe, ``Secrecy outage performance of transmit antenna selection for MIMO underlay cognitive radio systems over Nakagami-$m$ channels,'' \emph{IEEE Trans. Veh. Technol.}, vol. 66, no. 3, pp. 2237-2250, Mar. 2017.

\bibitem{Abramowitz1972}
M. Abramowitz and I. A. Stegun, \emph{Handbook of Mathematical Functions with Formulas, Graphs, and Mathematical Tables}, 9th. New York, NY, USA: Dover Press, 1972.





\end{thebibliography}
\end{document}